\documentclass[journal,twocolumn]{IEEEtran}
\usepackage{epsfig,color,amsmath,cite}
\usepackage{amsthm} 
\usepackage{amsmath}    
\usepackage[T1]{fontenc}
\usepackage[utf8]{inputenc}
\usepackage{bm}
\usepackage{epstopdf}
\usepackage{amssymb}
\usepackage{url}
\usepackage{enumitem} 
\usepackage{multirow}
\usepackage{hhline}
\usepackage{booktabs}
\usepackage{mathtools}
\usepackage{makecell}

\usepackage[linesnumbered,boxed,commentsnumbered,ruled,vlined,longend]{algorithm2e}
\makeatletter

\SetKwProg{Init}{Initialization}{}{Proceed}
\SetKwProg{St}{Start}{}{Proceed}
\makeatother
\usepackage{comment}

\makeatother
\DeclareMathAlphabet\mathbfcal{OMS}{cmsy}{b}{n}

\newtheorem{mydef}{Definition}

\usepackage{stackengine}

\newcommand{\mat}[1]{\boldsymbol{#1}}
\renewcommand{\vec}[1]{\boldsymbol{\mathrm{#1}}}

\providecommand{\mA}{\ensuremath{\mat{A}}}
\providecommand{\mB}{\ensuremath{\mat{B}}}
\providecommand{\mC}{\ensuremath{\mat{C}}}
\providecommand{\mD}{\ensuremath{\mat{D}}}
\providecommand{\mE}{\ensuremath{\mat{E}}}

\providecommand{\mQ}{\ensuremath{\mat{Q}}}
\providecommand{\mR}{\ensuremath{\mat{R}}}

\providecommand{\mT}{\ensuremath{\mat{T}}}

\providecommand{\mW}{\ensuremath{\mat{W}}}

\providecommand{\va}{\ensuremath{\vec{a}}}

\providecommand{\vf}{\ensuremath{\vec{f}}}

\providecommand{\vq}{\ensuremath{\vec{q}}}

\providecommand{\vu}{\ensuremath{\vec{u}}}

\providecommand{\vx}{\ensuremath{\vec{x}}}
\providecommand{\vy}{\ensuremath{\vec{y}}}

\newcommand{\m}{\boldsymbol}
\allowdisplaybreaks[4]
\usepackage[colorlinks = true,
linkcolor = blue,
urlcolor  = blue,
citecolor = blue,
anchorcolor = blue]{hyperref}

\newcommand{\mr}[1]{\mathrm{#1}}
\usepackage[framemethod=TikZ]{mdframed}
\mdfdefinestyle{MyFrame}{%
	linecolor=black,
	outerlinewidth=1.25pt,
	roundcorner=1.25pt,
	innerrightmargin=5pt,
	innerleftmargin=5pt,}
	
\usepackage{tikz}
\usetikzlibrary{matrix,positioning,decorations.pathreplacing}
\usetikzlibrary{shapes.geometric, arrows}
\usetikzlibrary{backgrounds}
\usetikzlibrary{shapes}
\usetikzlibrary{tikzmark}
\usetikzlibrary{calc}
\usetikzlibrary{arrows,shapes,positioning,shadows,trees,mindmap}
\usepackage[edges]{forest}
\usetikzlibrary{arrows.meta}
\colorlet{linecol}{black!75}
\usepackage{xkcdcolors} 

\usepackage[noabbrev]{cleveref}

\usepackage{mathtools}

\DeclarePairedDelimiter\abs{\lvert}{\rvert}%
\DeclarePairedDelimiter\norm{\lVert}{\rVert}%

\makeatletter
\let\oldabs\abs
\def\abs{\@ifstar{\oldabs}{\oldabs*}}
\let\oldnorm\norm
\def\norm{\@ifstar{\oldnorm}{\oldnorm*}}
\makeatother

\usepackage[english]{babel}
\usepackage[utf8]{inputenc}
\usepackage[super]{nth}

\usepackage{graphicx}
\usepackage{float}
\usepackage[caption = false]{subfig}

\usepackage{array}
\usepackage{threeparttable}

\usepackage[english]{babel}
\usepackage[utf8]{inputenc}
\usepackage[super]{nth}

\SetKwRepeat{Do}{do}{while}%

\everymath{\displaystyle}

\newcolumntype{E}{>{\centering\arraybackslash}m{0.5in}}
\newcolumntype{Q}{>{\centering\arraybackslash}m{3in}}
\newcolumntype{K}{>{\centering\arraybackslash}m{1.6in}}

\definecolor{airforceblue}{rgb}{0.36, 0.54, 0.66}
\usepackage[most]{tcolorbox}
\newtcbox{\colorboxouline}[1][]{boxsep=0pt,top=2.5pt,bottom=1pt,left=2pt,right=1pt,colframe=airforceblue,colback=airforceblue!10,boxrule=0pt,toprule=0pt,leftrule=3pt,sharp corners, enhanced jigsaw,,#1}

\usepackage[export]{adjustbox}
\usepackage{threeparttable}
\usepackage{multirow}

\usepackage{colortbl} 

\usepackage{mdframed}

\mdfdefinestyle{theoremstyle}{%
	linecolor=blue,linewidth=1pt,%
	 frametitlerule=true,%
	 frametitlebackgroundcolor=gray!20, innertopmargin=\topskip,}
	 
\usepackage[framemethod=TikZ]{mdframed}
\mdfsetup{skipabove=1mm,skipbelow=1mm}
\newcounter{theo}[section]

 	\definecolor{pansypurple}{rgb}{0.47, 0.09, 0.29}
 	\definecolor{darkcerulean}{rgb}{0.15, 0.38, 0.61}
 	\mdfdefinestyle{RepStyle}{%
 		skipabove=\topskip,
 		skipbelow=\topskip,
		innerleftmargin=0ex,
		innerrightmargin=0ex,
		innertopmargin=0ex,
 		linewidth=1.5pt,
 		leftline=false,
 		rightline=false,
 		frametitlefont={\color{darkcerulean}\normalfont\bfseries},
 		frametitlerule=false,
 		frametitlebackgroundcolor=white,
		backgroundcolor=white,
 		linecolor=darkcerulean,
 		theoremseparator={},
 		ntheorem=false 
 		} 	

\newmdenv[style=RepStyle]{Rep}

\definecolor{pinegreen}{rgb}{0.0, 0.47, 0.44}

\definecolor{cadmiumgreen}{rgb}{0.0, 0.42, 0.24}

\definecolor{persimmon}{rgb}{0.93, 0.35, 0.0}

\begin{document}

\title{\Large \textbf{Disinfectant Control in Drinking Water Networks: Integrating Advection-Dispersion-Reaction Models and Byproduct Constraints}}
	\author{Salma M. Elsheri$\text{f}^{\dagger, \P}$, Ahmad F. Tah$\text{a}^{\dagger, \ast \ast}$, and Ahmed A. Abokif$\text{a}^{\diamond}$
		\vspace{-0.7cm}
		\thanks{$^\dagger$Department of Civil and Environmental Engineering, Vanderbilt University, Nashville, TN, USA. Emails: salma.m.elsherif@vanderbilt.edu, ahmad.taha@vanderbilt.edu}
		\thanks{$\P$Secondary appointment: Department of Irrigation and Hydraulics Engineering, Faculty of Engineering, Cairo University.}
		\thanks{$^{\diamond}$Department of Civil, Materials, and Environmental Engineering, The University of Illinois Chicago. Email: abokifa@uic.edu}
		\thanks{$^{\ast \ast}$Corresponding author.}
		\thanks{This work is supported by the National Science Foundation under Grants 2151392 and 2015603.}
		}
	 
	\maketitle

	\begin{abstract}
		Effective disinfection is essential for maintaining water quality standards in distribution networks. Chlorination, as the most used technique, ensures safe water by maintaining sufficient chlorine residuals but also leads to the formation of disinfection byproducts (DBPs). These DBPs pose health risks, highlighting the need for chlorine injection control (CIC) by booster stations to balance safety and DBPs formation. Prior studies have followed various approaches to address this research problem. However, most of these studies overlook the changing flow conditions and their influence on the evolution of the chlorine and DBPs concentrations by integrating simplified transport-reaction models into CIC. In contrast, this paper proposes a novel CIC method that: (i) integrates multi-species dynamics, (ii) allows for a more accurate representation of the reaction dynamics of chlorine, other substances, and the resulting DBPs formation, and (iii) optimizes for the regulation of chlorine concentrations subject to EPA mandates thereby mitigating network-wide DBPs formation. The novelty of this study lies in its incorporation of time-dependent controllability analysis that captures the control coverage of each booster station. The effectiveness of the proposed CIC method is demonstrated through its application and validation via numerical case studies on different water networks with varying scales, initial conditions, and parameters.
	\end{abstract}
		
	\begin{IEEEkeywords}
		Disinfection Byproducts, Water Quality Control, Chlorine, Advection-Dispersion-Reaction, Booster Stations Injections
	\end{IEEEkeywords}

\markboth{In press, Journal of Water Research}{} 

\vspace{-0.3cm}
\section{\large Introduction}~\label{sec:Into-Lit}
\vspace{-0.2cm}
\IEEEPARstart{D}{infactants} play a pivotal role in water distribution networks (WDNs) to ensure compliance with water quality (WQ) safety standards. By maintaining the water pathogen-free, the occurrence of multiple waterborne diseases (e.g., cholera, typhoid) in the United States has significantly decreased since the utilization of disinfection process at the beginning of the $20^\text{th}$ century \cite{constableCenturyInnovationTwenty2003}. 
Of the various disinfection methods available, chlorination is the most commonly used technique in WDNs \cite{graymanwaltermHistoryWaterQuality2018}. While serving as a proxy for WQ monitoring, the objective is to have a sufficient chlorine residual all over the network to maintain safe water. However, chlorine actively reacts with various substances such as bacteria, organic matter, and microbial chemicals, leading to the formation of disinfection byproducts (DBPs) \cite{wangDisinfectionByproductFormation2013,zhouStabilityDrinkingWater2023,shah2024recent}. According to the Centers for Disease Control, exposure to these DBPs may increase the risk of cancer, liver damage, and decreased nervous system activity \cite{DisinfectionByproductsDBPs2022,kalita2024assessing}. Therefore, it's crucial that chlorine injections into the network by distributed booster stations are strategically controlled to achieve the primary goal of maintaining sufficient residuals and limiting the formation of DBPs.

Furthermore, accurate representation and detailed modeling are essential for effectively controlling of chlorine concentrations in water distribution networks. The transport and reaction dynamics of chlorine are modeled using advection-dispersion-reaction partial differential equations (ADR-PDEs). While some studies neglect the dispersion process, assuming that advection dominates in zones with high velocities, this overlooks the importance of dispersive transport in low-flow velocity pipes, especially in networks with laminar flow conditions and numerous dead-ends. This is because exclusively advective transport models may either underestimate or overestimate actual concentrations within the low-velocity zones \cite{abokifaWaterQualityModeling2016}. Additionally, in contrast to the widely adopted single-species chlorine decay model with a constant decay rate, multi-species models offer a more accurate representation of chlorine dynamics in WDNs \cite{jonkergouwVariableRateCoefficient2009, elsherifControltheoreticModelingMultispecies2023}. These models consider scenarios where chlorine interacts with other substances at rates different from the decay rate. Such scenarios include, but not limited, to contamination intrusion and substances derived from pipe materials. Furthermore, these multi-species models can be utilized to reflect the formation of DBPs, such as trihalomethanes (THMs), providing a comprehensive understanding of the WQ dynamics in WDNs \cite{moeiniBayesianOptimizationBooster2023}. 

To that end, this paper introduces a chlorine control approach that incorporates multi-species dynamics to maintain chlorine concentrations within EPA-specified limits \cite{acrylamideNationalPrimaryDrinking2009} while limiting the formation of DBPs. This control approach is formulated based on a detailed transport model that accounts for the dispersion effect in low-velocities zones of the network providing precise simulations of chlorine and other substances. Furthermore, the control approach \textit{strategically} allocates injections between booster stations based on a prior controllability analysis. This analysis implicitly prioritizes booster stations with broader coverage and assigns lower weights to zones with high existing DBPs concentrations. In the following section, we survey the literature on this topic and identify gaps that our study aims to address.

\vspace{-0.3cm}
\subsection{Literature Review}~\label{sec:lit}
The literature on chlorine modeling and control is broad and abundant, covering various reaction and decay dynamics, modeling techniques, and control approaches. Each of these aspects comes with its own set of underlying assumptions and limitations, contributing to the complexity of the topic. Next, we provide a brief summary of the literature on \textit{(i)} modeling the transport, decay, and reaction dynamics of chlorine and the subsequent formation of the DBPs, and \textit{(ii)} control techniques, objectives, constraints, and the specific scenarios under which they are applied.

\colorboxouline{\textbf{Chlorine Modeling and DBPs Formation}}
Accurate chlorine modeling is essential for the development of effective WQ control frameworks. Chlorine evolution in WDNs is covered by the transport, decay and reaction dynamics. The governing ADR-PDE divides these dynamics into three processes. The advection process accounts for the transport of chlorine along with the flowing water through the network, accounting for the movement propelled by the velocity of the water flow. The dispersion process captures the spreading of chlorine within the water flow caused by the irregularities and turbulence in the flow pattern and the diffusion of chlorine molecules across concentration gradients. Lastly, the reaction process includes chlorine decay, the formation of DBPs, and chlorine interaction with other substances in the water.

In WDNs, flow conditions are dependent on the consumers demand, connecting components, and pipes characteristics (i.e., diameter, length, material). Consequently, certain network zones exhibit turbulent flow conditions, while others demonstrate laminar flow conditions, particularly in dead-end sections. The transition between turbulent and laminar flow in various zones fluctuates throughout the day in response to changes in demand and the dynamics of storage components. In turbulent flow conditions and high flow velocities, advection has a greater influence than dispersion on the change of chlorine concentrations. Hence, in several studies \cite{graymanwaltermHistoryWaterQuality2018,engineeringconsultantImprovedWaterDistribution2012,elsherifControltheoreticModelingMultispecies2023} the dispersion process is neglected, with the focus placed on advection and reaction dynamics under the assumption of limited dead-end branches, higher velocities, and changing demands leading to a turbulent flow states. On the other hand, for WDNs with dead-ends comprising high percentage of the network, studies have proven that neglecting the dispersion effect results in inaccurate chlorine concentrations
\cite{tzatchkovAdvectionDispersionReactionModelingWater2002a, abokifaWaterQualityModeling2016, liImportanceDispersionNetwork2005}. Furthermore, authors in \cite{liImportanceDispersionNetwork2005} have linked the input patterns of chlorine injections to the dispersion effect, demonstrating the importance of including the dispersion in laminar flow zones when these injection patterns exhibit dramatic changes, a scenario common in many chlorine control applications. To that end, to build a generalized model for chlorine evolution in WDNs, the model has to be adaptable to various flow conditions and network characteristics, including layout and topology. This adaptability allows for the accurate simulation of the physically-representative dynamics. This generalized model to be integrated into chlorine control framework is a gap to be filled in this paper. 



In addition, a plethora of studies have investigated the modeling of the decay and reaction dynamics of chlorine in WDNs. These distinguish between these model vary in their order and the substances included \cite{helblingModelingResidualChlorine2009a}. The most commonly used model in developing chlorine control frameworks is the first-order single-species decay model. This model assumes that chlorine is linearly decaying in time as a result of its reaction with the bacteria and microbial organics in the water \cite{jonkergouwVariableRateCoefficient2009}. Nonetheless, this model fails to account for various scenarios where chlorine is actively reacting with other substances (e.g., contamination intrusion, substance from pipe's material), as well as the formation of DBPs \cite{fisherComprehensiveBulkChlorine2017a}. Conversely, the multi-species reaction model is an advanced approach capable of capturing these scenarios by introducing nonlinear reaction expressions that enable modeling the evolution of other substance(s) reacting with chlorine and the resultant formation of DBPs \cite{boccelliOptimalSchedulingBooster1998,elsherifControltheoreticModelingMultispecies2023,moeiniBayesianOptimizationBooster2023,liDisinfectantResidualStability2019a}. In this paper, we develop a chlorine control approach based on multi-species reaction model. This model simulates chlorine decay and reaction with a fictitious reactant, which can represent various substances or chemicals in a generalized form, and the resultant formation of DBPs.

Lastly, by defining the processes and reaction dynamics to be modeled by the ADR-PDEs, the techniques used to solve these equations differ. Due to the interplay between the dynamics of different components and the complexity added by the unique layout of each network, analytical solutions do not exist for these equations. Therefore, spatio-temporal discretization schemes are utilized to solve these equations and obtain the concentrations of chemicals \cite{rossmanNumericalMethodsModeling1996}. These schemes are categorized based on how the spatio-temporal grid is constructed and the interdependency between the calculations of grid points' concentrations. They are classified as Eulerian, Lagrangian, or hybrid Eulerian-Lagrangian based according to the grid construction approach. Eulerian schemes divide the grid into fixed-size meshes over space and time, while Lagrangian schemes use variable-sized segments. Hybrid schemes combine aspects of both, with fixed-size segments over time and variable segments over space, or vice versa. Moreover, they are classified based on which neighboring points' concentrations are considered in the calculation of a segment's concentration and whether these concentrations are from the previous time-step or the current one \cite{bashaEulerianLagrangianMethod2007a,rossmanNumericalMethodsModeling1996}. 

Several studies \cite{munavalliMULTISTEPEULERIANMETHOD2005,blokkerImportanceDemandModelling2007} employ one or more of these schemes in their studies of modeling the chlorine decay under advection-dominant conditions. Study \cite{elsherifControltheoreticModelingMultispecies2023} surveys the applicability of a range of Eulerian and Lagrangian based schemes on the chlorine multi-species advection-reaction PDEs, validating their performance against EPANET and its extension EPANET-MSX, which utilizes a Lagrangian-based scheme \cite{rossmanEPANETUserManual2020}. Other studies \cite{abokifaInvestigatingImpactsWater2020a,abokifaWaterQualityModeling2016,tzatchkovAdvectionDispersionReactionModelingWater2002a,ozdemirDiscussionLagrangianMethod2023} apply different schemes to simulate the chlorine decay and transport, while accounting for the dispersion effect. However, none of these studies investigate these scheme's performance while considering the multi-species dynamics and the alternation between advection- and dispersion-dominant conditions. 


%
%
%

\colorboxouline{\textbf{Chlorine Control Approaches and Underlying Scenarios}}

The topic of chlorine control is addressed in many studies, each with different objective function(s), constraints, and approaches to solving the problem, while falling within specific scenario frameworks \cite{fisherFrameworkOptimizingChlorine2018,ding2024application}. Several studies focus on minimizing chlorine injections while maintaining specific chlorine residual levels throughout the network, utilizing various optimization techniques such as linear programming \cite{boccelliOptimalSchedulingBooster1998,constansUsingLinearPrograms2012}, nonlinear programming \cite{mala-jetmarovaLostOptimisationWater2017}, and nonlinear genetic algorithm \cite{munavalliOptimalSchedulingMultiple2003}. Additionally, other studies couple the objective of minimization of chlorine injections with the booster stations allocation as multi-objective programming problem \cite{trybyFacilityLocationModel2002,pinedasandovalOptimalPlacementOperation2021a,ayvazIdentificationBestBooster2015}. 
Moreover, several studies have incorporated the DBPs formation issue into their optimization frameworks \cite{zhangOptimizingDisinfectantResidual2021}. A straightforward approach involves implicitly minimizing chlorine injections to reduce DBP formation \cite{prasadBoosterDisinfectionWater2004}. Furthermore, studies such as \cite{ardeshirControlTHMFormation2011,behzadianNovelApproachWater2012} tackle multi-objective optimization problems aiming to minimize both chlorine injections and THM formation, a specific byproduct of chlorine reactions. The authors in \cite{cozzolinoControlDBPsWater2005} investigate the optimal number and allocation of booster stations and chlorine injections to control DBPs formation under feasible operating scenarios. Moreover, Maheshwari et al. \cite{maheshwariOptimizationDisinfectantDosage2020} have addressed the issue in a multi-objective genetic optimization in an integrated MATLAB-EPANET-MSX platform, utilizing WQ model simulating chlorine residuals and DBPs levels.

We note a prevalent limitation across most of the aforementioned studies: they lack explicit articulation or formulation of WQ models that accurately represent the input/output relationship between booster stations injections and the concentrations at critical network components. Such WQ models facilitate the integration of state-of-the-art control algorithms within a control-theoretic framework. Efforts have been made to address this limitation by developing input/output WQ models and incorporating them into control frameworks \cite{shangParticleBacktrackingAlgorithm2002,wangAdaptiveControlWater2005,duzinkiewiczHierarchicalModelPredictive2005a}. Furthermore, Wang et al. \cite{wangHowEffectiveModel2021b} introduce a novel approach to model WQ dynamics using a linear state-space representation, enabling the modeling of system inputs, outputs, and chemical concentration states as a closed-form control-theoretic model. This study has combined this representation with a model predictive control (MPC) algorithm to determine chlorine injections into the system. However, this formulation is based solely on first-order single-species chlorine decay and transport dynamics. Additionally, this study neglects the dispersion process and employs an Eulerian discretization scheme that does not accurately represent advection-dominant processes. The study in \cite{elsherifComprehensiveFrameworkControlling2024} has expanded upon this control framework to address scenarios where substances interact with chlorine by utilizing the control-oriented nonlinear multi-species reaction model built on validated Eulerian discretization schemes \cite{elsherifControltheoreticModelingMultispecies2023} and has applied model order reduction techniques to manage the associated high dimensionality. However, this study does not consider varying flow conditions in the system or their impact on chemical transport dynamics. Furthermore, the control approach in \cite{elsherifComprehensiveFrameworkControlling2024} focuses on setting chlorine residuals within standardized limits, without considering DBPs formation or how network characteristics and flow conditions influence this formation and the control actions. Our paper aims to address these gaps, as detailed in the next section on our paper's contributions.

\subsection{Paper Contributions and Organization}
\vspace{0.2cm}
This paper’s main objective is to develop a comprehensive strategic WQ control approach that takes into account the different flow conditions across various zones within the network and their influence on the chemicals evolution, the complex dynamics of chlorine reaction and decay involving multiple species, and the formation of DBPs, which pose potential health risks. The development of this approach relies on a detailed analysis of system controllability under all these aspects, strategically guiding the control algorithm by weighting of input injections by booster stations to achieve enhanced controllability tailored to the unique characteristics of the network and hydraulics. Moreover, the approach seeks to mitigate the excessive formation of DBPs, thereby ensuring the maintenance of WQ standards and protecting public health. The corresponding detailed paper's contributions are as follows.
\begin{itemize}
	\item This paper extends and utilizes a multi-species reaction model that simulates chlorine decay and reaction with various substances, including the formation of DBPs. Moreover, this model accounts for the different hydraulic dynamics and flow conditions, capturing the advection- and dispersion-dominated states across different zones within the network. It dynamically switches between these states in certain zones while maintaining a single dominant mode throughout the simulation period in others. Formulated as a control-oriented state-space representation, this model integrates numerical discretization schemes to solve the ADR-PDE. The educational and theoretical significance of this work lies in its exploration of these schemes when coupled with multi-species dynamics and their impact on simulation variables such as time-step, discretization grid size, and numerical stability. This approach facilitates the integration of the model into control strategies while accommodating variations in process dynamics.
	\item The paper introduces a controllability analysis technique, providing valuable foresight into shaping the control strategy and optimizing the allocation of chlorine injections among booster stations. By leveraging insights from system controllability, the approach enhances the efficacy of control algorithms, enabling proactive decision-making and resource allocation to mitigate DBPs formation while ensuring WQ standards are met.
	\item  Development of a control approach that integrates detailed modeling insights with real-world operational considerations. This approach offers a generalized solution applicable to diverse network configurations and hydraulic settings, accommodating predetermined booster station locations and evolving network dynamics. Through thorough validation on different networks and case studies, the proposed control approach demonstrates its effectiveness in addressing the dynamic challenges of WQ regulation.
\end{itemize}

The remainder of this paper is structured as follows: Section \ref{sec:MSDynModel} presents the multi-species disinfectant and byproducts dynamics model, detailing the governing equations governing the evolution of chemicals across network components. Based on this model, the formulation of the disinfectant control problem is derived in Section \ref{sec:CLDBPCntrlProbForm}. In Section \ref{sec:CaseStudies}, we validate our proposed approach through several case studies, encompassing various scales, layouts, and scenarios. Finally, Section \ref{sec:ConcLimRec} offers conclusions, discusses the limitations of the study, and provides recommendations for future research directions.

\pagebreak
\textbf{Notation.} Italicized, boldface upper and lower case characters represent matrices and column vectors: $a$ is a scalar, $\va$ is a vector, and $\mA$ is a matrix. The notation $\mathbb{R}^n$ denotes the sets of column vectors with $n$ real numbers, while $\mathbb{R}^{n \times m}$ denotes the sets of matrices with $n$ rows and $m$ columns. Variables with upper case characters $\boldsymbol{\cdot}^\mathrm{J}, \boldsymbol{\cdot}^\mathrm{R}, \boldsymbol{\cdot}^\mathrm{TK}, \boldsymbol{\cdot}^\mathrm{P}, \boldsymbol{\cdot}^\mathrm{M},$ and $\boldsymbol{\cdot}^\mathrm{V}$ correspond to junctions, reservoirs, tanks, pipes, pumps, and valves, respectively. Additionally, $\boldsymbol{\cdot}^\mathrm{B_\mathrm{TK}}$ and $\boldsymbol{\cdot}^\mathrm{B_\mathrm{J}}$ represent variables associated with booster stations located at tanks and junctions, while $\boldsymbol{\cdot}^\mathrm{D_\mathrm{J}}$ donates the demand variables at the junctions.

\section{Multi-species Disinfectant and Byproducts Dynamics Model}~\label{sec:MSDynModel}

We model the WDN by a directed graph $\mathcal{G} = (\mathcal{N},\mathcal{L})$.  The set $\mathcal{N}$ defines the nodes and is partitioned as $\mathcal{N} = \mathcal{J} \cup \mathcal{T} \cup \mathcal{R}$ where sets $\mathcal{J}$, $\mathcal{T}$, and $\mathcal{R}$ are collections of junctions, tanks, and reservoirs. Let $\mathcal{L} \subseteq \mathcal{N} \times \mathcal{N}$ be the set of links, and define the partition $\mathcal{L} = \mathcal{P} \cup \mathcal{M} \cup \mathcal{V}$, where sets $\mathcal{P}$, $\mathcal{M}$, and $\mathcal{V}$ represent the collection of pipes, pumps, and valves. Total number of states is $n_x=n_L+n_N$, where $n_\mathrm{L}$ and $n_\mathrm{N}$ are numbers of links and nodes. The number of reservoirs, junctions, tanks, pumps, valves, and pipes are $n_\mathrm{R}, n_\mathrm{J}, n_\mathrm{TK}, n_\mathrm{M}, n_\mathrm{V},$ and $n_\mathrm{P}$. Each pipe $i$ with length $L_i$ is spatially discretized and split into $s_{L_i}$ segments. Hence, the number of links states is expressed as $n_\mathrm{L}=n_\mathrm{M}+n_\mathrm{V}+ \sum_{i=1}^{n_\mathrm{P}} s_{L_i}$ while $n_\mathrm{N}=n_\mathrm{R}+n_\mathrm{J}+n_\mathrm{TK}$ represents the number of nodes states.

We apply the principles of conservation of mass, transport, decay, and reaction to model the evolution of chemicals and substances throughout the different components of the WDNs. This modeling assumes that the hydraulic variables and settings are pre-determined and operate under non-transient conditions for extended period simulations. This means that the network operates on a steady-state basis for each time step, assuming that the system reaches equilibrium within each interval before moving to the next. In the following sections, we provide the governing equations for the transport of any chemical by denoting the concentration by $c$. Additionally, for dynamics that depend on specific chemical properties, such as chlorine decay and mutual reactions with other substances, we specify which chemical's concentration we are calculating by adding a special notation to the symbol $c$.  

\subsection{Transport and Reaction in Pipes: Advection-Dispersion-Reaction Model}~\label{sec:TranReactPipes}
In our paper, we simulate the transport and reaction in pipes by the ADR-PDE, which for Pipe $i$ is expressed as
	\begin{equation}\label{equ:PDE}
	\partial_t c_i^\mathrm{P}=-{v_i(t)} \partial_x c_i^\mathrm{P} + D_i(t) \partial_{xx} c_i^\mathrm{P} + R^\mathrm{P}(c_i^\mathrm{P}(x,t)),
\end{equation}
where $c^\mathrm{P}_i(x,t)$ is concentration in pipe at location $x$ along its length and time $t$; $v_i(t)$ is the mean flow velocity; $D_i(t)$ is the effective longitudinal dispersion coefficient of the chemical; and $R^\mathrm{P}(c^\mathrm{P}_i(x,t))$ is the decay and reaction expression in pipes (more explanation is given in Section \ref{sec:MSmodel}). 

In Eq. \eqref{equ:PDE}, the term ${v_i(t)} \partial_x c_i^\mathrm{P}$ describes the change in concentration as an impact of the advection process. Advection causes translation of the chemical by moving it with the flow velocity. Whilst, the term $D_i(t) \partial_{xx} c_i^\mathrm{P}$ represents the impact of chemical dispersion, where the effective longitudinal dispersion coefficient $D_i(t)$ encompasses both molecular diffusion and shear dispersion caused by the unevenness of the velocity profile. The calculation of $D_i(t)$ for Pipe $i$ is dependent on the flow condition, which can be categorized as laminar, transitional, or turbulent flow based on the unitless Reynolds number $Re_i$
\begin{equation}~\label{eq:ReynoldsNum}
	Re_i(t) = \frac{\rho L_{D_i}}{\mu}v_i(t),
\end{equation}
where $\rho$ represents the water density, which is typically 62.4 lb/ft$^3$ (998.4 kg/m$^3$), $\mu$ denotes the water dynamic viscosity, typically $2.42 \times 10^{-5}$ lb$\cdot$ft$^{-1}\cdot$s$^{-1}$ ($3.6 \times 10^{-5}$ kg$\cdot$m$^{-1}\cdot$s$^{-1}$) at room temperature, and $L_{D_i}$ stands for the characteristic length, which is the hydraulic diameter for water flowing pressurized in a pipe, equal to the pipe diameter $d_{\mathrm{P}_i}$.

For laminar flow conditions ($Re < 2300$), $D_i(t)$ is calculated as an averaged value over the solute residence time \cite{leeMassDispersionIntermittent2004}:
\begin{equation}~\label{eq:DispCoefLam}
	D_i(t) = \frac{{d^2_{\mathrm{P}_i}} {v^2_i(t)}}{12 d_\mathrm{m}} \Bigg[ 1- \Bigg[ \frac{1- \exp (- \frac{4d_\mathrm{m} t_r(t)}{d^2_{\mathrm{P}_i}})}{ \frac{4d_\mathrm{m} t_r(t)}{d^2_{\mathrm{P}_i}}} \Bigg] \Bigg],
\end{equation}
where $t_r$ is the pipe residence time, $t_r(t)=\frac{L_i}{v_i(t)}$; and $d_\mathrm{m}$ is the molecular diffusion coefficient---references \cite{leaist1986absorption,leeMassDispersionIntermittent2004,rossmanEPANETUserManual2020} provide values for this coefficient for chlorine and different by-products.

For turbulent and transitional flow conditions ($Re \geq 2300$), $D_i(t)$ does not depend on the molecular diffusion coefficient, rather on the flow condition through the pipe \cite{bashaEulerianLagrangianMethod2007a}:
\begin{equation}~\label{eq:DispCoefTurb}
	D_i(t) = \frac{d_{\mathrm{P}_i} v_{\tau_i}(t)}{2}  \Big[ 10.1 + 577 \Big( \frac{Re_i(t)}{100} \Big)^{-2.2} \Big],
\end{equation}
where $v_{\tau}$ ia the shear velocity taken as a percentage of the mean flow velocity.

%
%


The influence of dispersion may be minimal in certain sections or across the entire network when turbulent conditions prevail. In such cases, the advection process is the dominant process in the chemical transportation throughout these network segments. However, it is crucial to account for dispersion when the flow velocity is low and the flow approaches the laminar state (e.g., dead-end segments). The significance of dispersion can be assessed quantitatively using the Peclet number. For Pipe $i$, the Peclet number $Pe_i$ at a given time $t$ can be calculated using Eq. \eqref{eq:PecletNum}. The Peclet number serves as a dimensionless indicator of the relative importance between advection and dispersion, with a high value suggesting a flow scenario primarily governed by advection, where dispersion can be neglected. Certain simulation software establish a threshold value for the Peclet number $Pe^\mathrm{th}$ beyond which dispersion effect is neglected (e.g., EPANET sets this threshold at 1000 \cite{rossmanEPANETUserManual2020}).

\begin{equation}~\label{eq:PecletNum}
	Pe_i(t)=\frac{v_i(t) L_i}{D_i(t)}.
\end{equation}

Following the explanation of all parameters and variables in Eq. \eqref{equ:PDE}, the subsequent step is to discuss how it is actually solved. Notably, there is no analytical solution for Eq. \eqref{equ:PDE} in many scenarios with complex dynamics and different components models. That is, numerical discretization schemes are employed. These schemes discretize the pipe over a spacio-temporal grid, and accordingly concentrations are calculated at each segment of this grid. A variety of schemes have been developed and utilized in the literature \cite{rossmanNumericalMethodsModeling1996}. The choice of the scheme depends on the actual physics of the system and the processes involved in the simulation of chemicals evolution. For instance, if the dispersion influence is neglected, a different method should be chosen compared to scenarios where it is taken into consideration. To that end, we detail herein numerical schemes to be employed for both scenarios (i.e., ADR-PDE and AR-PDE). Afterwards, we explain how to determine the grid size and simulation time-step, while ensuring numerical stability. That is, for the first two sections (\ref{sec:AdvDisc} and \ref{sec:DispDisc}), we use the notation of $\Delta t$ and $\Delta x$ for the time-step and segment size for each pipe. Following, we introduce how to determine these parameters in Section \ref{sec:DtDxStab}.

\subsubsection{Advection-Dominant PDE Discretization Schemes}~\label{sec:AdvDisc}
In advection-dominant transport, the main two processes are the advection where the concentration at a certain location and time is affected by upstream concentrations, and reaction where chemicals decay and/or mutually react. That being said, \textit{Upwind} discretization schemes are more descriptive to the actual physical process considered among other schemes \cite{hirschNumericalComputationInternal1990}. In this paper we consider both \textit{Explicit} and \textit{Implicit} Upwind schemes to investigate their performance in the proposed control framework. The distinction between \textit{explicit} and \textit{implicit} notations is clarified in Definition \ref{def:ImpvsExp}.

\begin{mydef}~\label{def:ImpvsExp}	
	A discretization scheme is referred to as \textit{explicit} or \textit{implicit} depending on whether the concentrations of neighboring segments/nodes, on which the concentration of the segment being calculated depends, are taken from the previous time-step (and thus determined); Explicit, or from the current time-step (and thus solved for simultaneously); Implicit.
\end{mydef}

For Pipe $i$ divided into $n_{\mathrm{s}_i}$ segments, the concentrations at $t+\Delta t$ of the first, last, and any in-between segment (i.e., $c_i^\mathrm{P}(1,t +\Delta t)$, $c_i^\mathrm{P}(s,t +\Delta t)$, $c_i^\mathrm{P}(s_\mathrm{L},t +\Delta t)$) is calculated as listed in Tab. \ref{tab:DiscSchms} for both schemes. In these calculations, $\tilde{\lambda}_i(t)$ is the Courant number (CN) and expressed as
\begin{equation}~\label{eq:CourantNum}
	\tilde{\lambda}_i(t)=\frac{v_i(t) \Delta t}{\Delta x_i}.
\end{equation}

\subsubsection{Dispersion-Dominant PDE Discretization Schemes}~\label{sec:DispDisc}
By considering the dispersion process in modeling the transport of chemicals, both upstream and downstream concentrations influence the concentration at a specific location. Thus, we employ two central Eulerian discretization schemes to solve the ADR-PDE: the explicit Lax-Wendroff (L-W) Scheme and the implicit Backward Euler Scheme. Refer to Definition \ref{def:ImpvsExp} for differentiation between explicit and implicit schemes. Similar to the previous schemes, Tab. \ref{tab:DiscSchms} provides the formulations for the concentrations at $t+\Delta t$ of the first, last, and any in-between segment (i.e., $c_i^\mathrm{P}(1,t +\Delta t)$, $c_i^\mathrm{P}(s,t +\Delta t)$, $c_i^\mathrm{P}(s_\mathrm{L},t +\Delta t)$) of Pipe $i$ by utilizing these schemes. The coefficients used in these formulations are the dispersion number expressed as
\begin{equation}~\label{eq:DispNum}
	{\alpha}_i(t)=D_i(t) \frac{\Delta t}{\Delta x_i^2}, 
\end{equation} 
while $\underline{\lambda}_i(t), \lambda_i(t),$ and $\overline{\lambda}_i(t)$ are the weighting coefficients calculated as follows
\begin{equation}~\label{equ:LW_coeff}
	\begin{aligned}
		\underline{\lambda}_i(t) &= 0.5\tilde{\lambda}_i(t)\left(1+{\tilde{\lambda}}_i(t)\right), \\
		\lambda_i(t) &=  1-{\tilde{\lambda}}_i^2(t),  \\
		\overline{\lambda}_i(t) &= -0.5\tilde{\lambda}_i(t)\left(1-\tilde{\lambda}_i(t)\right).
	\end{aligned}
\end{equation}

\bgroup
\setlength\tabcolsep{1.5pt}
\begin{table*}[t!]
	\centering
	\caption{Expressions for four discretization schemes; two applied for the AR-PDE: \textcolor{darkcerulean}{Explicit Upwind} ({EU}) and \textcolor{darkcerulean}{Implicit Upwind} ({IU}), and the other two are for solving the ADR-PDE: \textcolor{pansypurple}{Lax-Wendroff} ({L-W}) and \textcolor{pansypurple}{Backward Euler} ({BE}). The header row of the table includes variables representing the concentrations at the upstream Junction $j$, first, intermediary, last segments of Pipe $i$, and the downstream Junction $k$ at times $t$ and $t+ \Delta t$. For each scheme, three expression can be formulated from this table depending on the variables in the header row: these expressions are for the first, any intermediary, and last segments of the pipe. To formulate these expressions, treat each row independently. Multiply the values in each cell by the corresponding header variable of the same column. Then, sum these resultant terms on each side of the equal sign to complete the expression. In the column labeled $R^\mathrm{P}_{\mathrm{MS}}(\textcolor{pinegreen}{\cdot}) \Delta t$, the cells specify the variable upon which this function is calculated for. Note that, for simplicity and a more compact table, we eliminate writing the time factor for all the parameters $\alpha, \tilde{\lambda}, \lambda, \underline{\lambda},$ and $\overline{\lambda}$ of Eq. \eqref{eq:CourantNum}, \eqref{eq:DispNum}, and \eqref{equ:LW_coeff}, knowing that all of them are taken at time $(t)$. 	\vspace{-0.35cm}}~\label{tab:DiscSchms}
	\begin{tabular}{l|c|c|c|c|c||c||c|c|c|c|c|c||c|}
		\cline{2-14}                                                                  
		&  $c^\mathrm{J}_j(t+\Delta t)$                 & $c^\mathrm{P}_i(1,t+ \Delta t)$              & $c^\mathrm{P}_i(s,t+ \Delta t)$              & $c^\mathrm{P}_i(s_\mathrm{L},t+ \Delta t)$ & $c^\mathrm{J}_k(t+\Delta t)$      &        & $c^\mathrm{J}_j(t)$                       & $c^\mathrm{P}_i(1,t)$                     & $c^\mathrm{P}_i(s,t)$                     & $c^\mathrm{P}_i(s_\mathrm{L},t)$         & $c^\mathrm{J}_k(t)$                      & $R^\mathrm{P}_{\mathrm{MS}}(\textcolor{pinegreen}{\cdot}) \Delta t$ & Eq.      \\
		\cline{2-14}                                                                 \noalign{\vskip\doublerulesep
			\vskip-\arrayrulewidth}  \hline
		\multicolumn{1}{|l|}{\multirow{3}{*}{\textbf{\begin{tabular}[c]{@{}l@{}} \textcolor{darkcerulean}{Explicit} \\ \textcolor{darkcerulean}{Upwind} \\ \end{tabular}}}} &                                              & 1                                            &                                              &                                            &                                          & = & $\tilde{\lambda}_i$                          & $1-\tilde{\lambda}_i$                        &                                           &                                          &                                          & $(\textcolor{pinegreen}{c^\mathrm{P}_i(1,t)})$                      &  \begin{subequations}~\label{eq:EU_1}\hspace{-0.1cm}\end{subequations}\eqref{eq:EU_1}  \\ \cline{2-14} 
		\multicolumn{1}{|l|}{}                                                                                     &                                              &                                              & 1                                            &                                            &                & =                         &                                           & $\tilde{\lambda}_i$                          & 	$1-\tilde{\lambda}_i$                        &                                          &                                          & $(\textcolor{pinegreen}{c^\mathrm{P}_i(s,t)})$                      & \begin{subequations}~\label{eq:EU_s}\hspace{-0.1cm}\end{subequations}\eqref{eq:EU_s}   \\ \cline{2-14} 
		\multicolumn{1}{|l|}{}                                                                                     &                                              &                                              &                                              & 1                                          &              &  =                            &                                           &                                           & $\tilde{{\lambda}}_i$                          & $1-\tilde{\lambda}_i$                       &                                          & $(\textcolor{pinegreen}{c^\mathrm{P}_i(s_\mathrm{L},t)})$           & \begin{subequations}~\label{eq:EU_sL}\hspace{-0.1cm}\end{subequations}\eqref{eq:EU_sL} \\ \hline \hline
		\multicolumn{1}{|l|}{\multirow{3}{*}{\textbf{\begin{tabular}[c]{@{}l@{}} \textcolor{darkcerulean}{Implicit} \\ \textcolor{darkcerulean}{Upwind} \\ \end{tabular}}}} & $-\tilde{\lambda}_i$                            & $1+\tilde{\lambda}_i$                           &                                              &                                            &                                     & =     &                                           & 1                                         &                                           &                                          &                                          & $(\textcolor{pinegreen}{c^\mathrm{P}_i(1,t)})$                      & \begin{subequations}~\label{eq:IU_1}\hspace{-0.1cm}\end{subequations}\eqref{eq:IU_1}   \\ \cline{2-14} 
		\multicolumn{1}{|l|}{}                                                                                     &                                              & $-\tilde{\lambda}_i$                            & $1+\tilde{\lambda}_i$                           &                                            &                                 & =         &                                           &                                           & 1                                         &                                          &                                          & $(\textcolor{pinegreen}{c^\mathrm{P}_i(s,t)})$                      & \begin{subequations}~\label{eq:IU_s}\hspace{-0.1cm}\end{subequations}\eqref{eq:IU_s}   \\ \cline{2-14} 
		\multicolumn{1}{|l|}{}                                                                                     &                                              &                                              & $-\tilde{\lambda}_i$                            & $1+\tilde{\lambda}_i$                         &            & =                              &                                           &                                           &                                           & 1                                        &                                          & $(\textcolor{pinegreen}{c^\mathrm{P}_i(s_\mathrm{L},t)})$           & \begin{subequations}~\label{eq:IU_sL}\hspace{-0.1cm}\end{subequations}\eqref{eq:IU_sL} \\ \hline \hline
		\multicolumn{1}{|l|}{\multirow{3}{*}{\textbf{\begin{tabular}[c]{@{}l@{}} \textcolor{pansypurple}{Lax-}\\\textcolor{pansypurple}{Wendroff}\\ \end{tabular}}}}    &                                              & 1                                            &                                              &                                            &                            & =              & $\underline{\lambda}_i+\alpha_i$ & $\lambda_i-2\alpha_i$            & $\overline{\lambda}_i+\alpha_i$  &                                          &                                          & $(\textcolor{pinegreen}{c^\mathrm{P}_i(1,t)})$                      &   \begin{subequations}~\label{eq:LW_1}\hspace{-0.1cm}\end{subequations}\eqref{eq:LW_1}   \\ \cline{2-14} 
		\multicolumn{1}{|l|}{}                                                                                     &                                              &                                              & 1                                            &                                            &                      & =                    &                                           & $\underline{\lambda}_i+\alpha_i$ & $\lambda_i-2\alpha_i$            & $\overline{\lambda}_i+\alpha_i$ &                                          & $(\textcolor{pinegreen}{c^\mathrm{P}_i(s,t)})$                      &  \begin{subequations}~\label{eq:LW_s}\hspace{-0.1cm}\end{subequations}\eqref{eq:LW_s}   \\ \cline{2-14} 
		\multicolumn{1}{|l|}{}                                                                                     &                                              &                                              &                                              & 1                                          &                                & =          &                                           &                                           & $\underline{\lambda}_i+\alpha_i$ & $\lambda_i-2\alpha_i$           & $\overline{\lambda}_i+\alpha_i$ & $(\textcolor{pinegreen}{c^\mathrm{P}_i(s_\mathrm{L},t)})$           &   \begin{subequations}~\label{eq:LW_sL}\hspace{-0.1cm}\end{subequations}\eqref{eq:LW_sL}    \\ \hline \hline
		\multicolumn{1}{|l|}{\multirow{3}{*}{\textbf{\begin{tabular}[c]{@{}l@{}}\textcolor{pansypurple}{Backward}\\ \textcolor{pansypurple}{Euler}\\ \end{tabular}}}}   & $-0.5\tilde{\lambda}_i-\alpha_i$ & $1+2\alpha_i$                          & $0.5\tilde{\lambda}_i-\alpha_i$     &                                            &                                          &          = &                                 & 1                                         &                                           &                                          &                                          & $(\textcolor{pinegreen}{c^\mathrm{P}_i(1,t)})$                      &    \begin{subequations}~\label{eq:BE_1}\hspace{-0.1cm}\end{subequations}\eqref{eq:BE_1}     \\ \cline{2-14} 
		\multicolumn{1}{|l|}{}                                                                                     &                                              & $-0.5\tilde{\lambda}_i-\alpha_i$ & $1+2\alpha_i$                          & $0.5\tilde{\lambda}_i-\alpha_i$   &                                          &                    = &                        &                                           & 1                                         &                                          &                                          & $(\textcolor{pinegreen}{c^\mathrm{P}_i(s,t)})$                      &   \begin{subequations}~\label{eq:BE_s}\hspace{-0.1cm}\end{subequations}\eqref{eq:BE_s}   \\ \cline{2-14} 
		\multicolumn{1}{|l|}{}                                                                                     &                                              &                                              & $-0.5\tilde{\lambda}_i-\alpha_i$ & $1+2\alpha_i$                        & $0.5\tilde{\lambda}_i-\alpha_i$ &  =                                          &                              &              &                                           & 1                                        &                                          & $(\textcolor{pinegreen}{c^\mathrm{P}_i(s_\mathrm{L},t)})$           &     \begin{subequations}~\label{eq:BE_sL}\hspace{-0.1cm}\end{subequations}\eqref{eq:BE_sL}    \\ \hline
	\end{tabular}
\end{table*}
\egroup

{\setcounter{algocf}{0}
	\begin{algorithm}[t!]
		\small	\DontPrintSemicolon
		\SetAlgorithmName{Procedure}
		1 1\KwIn{WDN topology, components’ characteristics, hydraulics
			parameters, and $Pe^\mathrm{th}$}
		\KwOut{Time-step $\Delta t$ and number of segmenents and segment size for each pipe; $s_{\mathrm{L}_i}$ and $\Delta x_i,$ $i=1, \ldots, n_\mathrm{P}$  } 
			Initialize $\Delta t_\mathrm{temp} \leftarrow 0$ \;
			Initialize $\Delta t \leftarrow 10^6$ \textcolor{blue}{// Start with a large number} \;
			\For{$i = 1$ \textbf{to} $n_\mathrm{P}$}{
			Initialize $\mD_i \leftarrow \varnothing$ and $\boldsymbol{Pe}_i \leftarrow \varnothing$ \;
			\For{each hydraulic time-step}{
			Calculate $Re_i(t)$ via \eqref{eq:ReynoldsNum} \;
			\uIf{$Re_i(t) < 2300$}{Calculate $D_i(t)$ via \eqref{eq:DispCoefLam}}
			\Else{Calculate $D_i(t)$ via \eqref{eq:DispCoefTurb}}
			Calculate $Pe_i(t)$ via \eqref{eq:PecletNum}\;
			Append $D_i(t) \rightarrow \mD_i$ and $Pe_i(t) \rightarrow \boldsymbol{Pe}_i$ \;
			\uIf{$Pe_i(t) \leq Pe^\mathrm{th}$}{
			$\Delta t_\mathrm{temp} = {2D_i(t)}/{v_i^2(t)}$ \textcolor{blue}{// To avoid spatial oscillations}\;
			\uIf{$\Delta t_\mathrm{temp} < \Delta t$}{$\Delta t \leftarrow \Delta t_\mathrm{temp} $}{\textbf{end if}}}{\textbf{end if}}
			}
			}
			\textbf{return} $\Delta t$ \textcolor{blue}{// Final time-step} \;
			\uIf{Applying Explicit Discretization Methods}{
		\For{$i= 1$ \textbf{to} $n_\mathrm{P}$}{
			Initialize $\Delta x_{\mathrm{temp}_1} \leftarrow 0$ and $\Delta x_{\mathrm{temp}_2} \leftarrow 0$ \;
			Initialize $\Delta x \leftarrow 0$ \; 
		\For{each hydraulic time-step}{
			\uIf{$Pe_i(t) \leq Pe^\mathrm{th}$}{
			$\Delta x_{\mathrm{temp}_1} = ({2 D_i(t) \Delta t})^{-\frac{1}{2}}$ \textcolor{blue}{// von Neumann condition}}
			\Else{$\Delta x_{\mathrm{temp}_2} = v_i(t) \Delta t$ \textcolor{blue}{// Courant-Friedrichs-Lewy condition}}
			$\Delta x_i \leftarrow \max\{ \Delta x_{\mathrm{temp}_1}, \Delta x_{\mathrm{temp}_2}, \Delta x_i \}$ 
		}
		$s_{\mathrm{L}_i} = \lfloor L_i / \Delta x_i \rfloor$ \;
 		$\Delta x_i = L_i / s_{\mathrm{L}_i}$
		}
		}\ElseIf{Applying Implicit Discretization Methods}{
		Set $s_{\mathrm{L}_i}$ as fixed arbitrary integer for all $i=1, \ldots, n_\mathrm{P}$ \;
		$\Delta x_i = L_i / s_{\mathrm{L}_i}, \; \forall i=1, \ldots, n_\mathrm{P}$}
		\textbf{return} $s_{\mathrm{L}_i}$ and $\Delta x_i, \; \forall i=1, \ldots, n_\mathrm{P}$ \textcolor{blue}{// Final number of segments and segment size for each pipe}
		\caption{Time-step and grid size determination and numerical stability assurance\label{alg:Proc1}}
\end{algorithm}}

\subsubsection{Pre-Simulation Preparations: Time-Step, Grid Size, and Numerical Stability Assurance}~\label{sec:DtDxStab}
Applying explicit discretization methods requires satisfying numerical stability conditions. For the Explicit Upwind Scheme applied on AR-PDE, the Courant-Friedrichs-Lewy condition states that the CN number is to be maintained in the range of $0<\tilde{\lambda}_i(t) \leq 1$ for Pipe $i$. Additionally, to solve the ADR-PDE using L-W scheme for Pipe $i$ and ensure its numerical stability, the von Neumann condition states that $0 < {\tilde{\lambda}_i}^2(t) \leq 2 {\alpha_i}(t) \leq 1$. In addition, for any central scheme applied for ADR-PDE, the ratio $\tilde{\lambda}_i(t)/\tilde{\alpha}_i(t)$ is set to be less or equal to 2 to achieve high accuracy and avoid spatial oscillations \cite{durran2010numerical}.

That being said, different conditions need to be satisfied, according to which the time-step and grid size are determined. In Procedure \ref{alg:Proc1}, we list the steps needed to compute these parameters while following a numerically stable approach.

\subsubsection{Dynamic Modeling of the Transport and Reaction Dynamic in Pipes}\label{sec:DynamicPDEModeling}
As explained in the previous sections, our approach facilitates dynamic switching between discretization methods based on the prevailing processes in transport and reaction dynamics. 
As a starter, Procedure \ref{alg:Proc1} ensures consistency in system dimensions (specifically, the WQ time-step and number of segments for each pipe) while maintaining numerical stability throughout the simulation period, regardless of whether dispersion effects are considered. To that end, the switch between the discretization methods is solely about the considered elements and the calculations of the dependency between the segments and nodes from a time-step to the next (see Tab. \ref{tab:DiscSchms}). With each hydraulic time-step for updating system matrices, an additional condition is introduced to assess the current Peclet number (Eq. \eqref{eq:PecletNum}) against the threshold $Pe_{\text{th}}$, determining whether dispersion effects should be neglected. Accordingly, parameters and elements outlined in Tab. \ref{tab:DiscSchms} are computed.

\subsection{Conservation of Mass}
For nodes and links other than pipes, the conservation of mass principle is applied to formulate the governing equations for concentration calculations within water distribution networks.

\subsubsection{Mass Balance at Reservoirs} For any Reservoir $i$, concentration of a chemical  is assumed to remain constant over time, such that $c_i^\mathrm{R}(t + \Delta t) = c_i^\mathrm{R}(t)$.

\subsubsection{Mass Balance at Pumps and Valves} In our WQ model, we deal with pumps and valves as transmission links with negligible length. Accordingly, chemicals concentrations at these elements are taken equal to the concentration of the upstream node. That being said, for  Pump $i$ or Valve $j$ installed after Reservoir/Tank/Junction $k$, concentrations are expressed as $c_i^{\mathrm{M}}(t+\Delta t) = c_k^{\boldsymbol{\cdot}}(t+\Delta t),$ and $c_j^{\mathrm{V}}(t+\Delta t) = c_k^{\boldsymbol{\cdot}}(t+\Delta t).$

\begin{table*}[t!]
	\centering
	\caption{Chlorine multi-species reaction and decay dynamics and DBPs formation models. In these models, concentrations for chlorine, fictitious reactant, and THMs are donated by $c$, \textcolor{cadmiumgreen}{$\tilde{c}$}, and \textcolor{persimmon}{$\hat{c}$}, respectively. \vspace{-0.15cm}}~\label{tab:MSExp}
	\begin{tabular}{l|c|c|}
		\cline{2-3}
		& Segment $s$ of Pipe $i$                                                                                                  & Tank $j$                                                                                                               \\ \hline
		\multicolumn{1}{|l|}{$R_{\mathrm{MS}}(c(t))$}  & $-k^\mathrm{P}_i c^\mathrm{P}_i(s,t)-k_r c^\mathrm{P}_i(s,t) \textcolor{cadmiumgreen}{\tilde{c}^\mathrm{P}_i(s,t)}$ & $-k^\mathrm{TK}_j c^\mathrm{TK}_j(t)-k_r c^\mathrm{TK}_j(t) \textcolor{cadmiumgreen}{\tilde{c}^\mathrm{TK}_j(t)}$ \\ \hline
		\multicolumn{1}{|l|}{$R_{\mathrm{MS}}(\textcolor{cadmiumgreen}{\tilde{c}(t)})$}  & $- Y_\mathrm{FR} k_r c^\mathrm{P}_i(s,t) \textcolor{cadmiumgreen}{\tilde{c}^\mathrm{P}_i(s,t)}$                                  & $- Y_\mathrm{FR} k_r c^\mathrm{TK}_j(t) \textcolor{cadmiumgreen}{\tilde{c}^\mathrm{TK}_j(t)}$                                  \\ \hline
		\multicolumn{1}{|l|}{$R_{\mathrm{MS}}(\textcolor{persimmon}{\hat{c}(t)})$} & $ Y_\mathrm{THMs}  k_r c^\mathrm{P}_i(s,t) \textcolor{cadmiumgreen}{\tilde{c}^\mathrm{P}_i(s,t)}$                                  & $ Y_\mathrm{THMs} k_r c^\mathrm{TK}_j(t) \textcolor{cadmiumgreen}{\tilde{c}^\mathrm{TK}_j(t)}$                                  \\ \hline
	\end{tabular}
\end{table*}

\subsubsection{Mass Balance at Junctions} Water from all inflows into a junction is assumed to have complete and instantaneous mixing. That is, while assuming that there is no storage time at junctions, at a Junction $i$, all outflows have the same concentration for a specific chemical. This concentration is expressed as
\begin{equation}~\label{equ:mb-junc} 
	c_i^\mathrm{J}(t)= \frac{\sum_{j \in L_{\mathrm{in}}} q_{\mathrm{in}}^{j}(t) c_\mathrm{in}^j(t)+q^\mathrm{B_\mathrm{J}}_i(t) c^\mathrm{B_\mathrm{J}}_i(t)}{q^{\mathrm{D}_\mathrm{J}}_i(t)+\sum_{k \in L_{\mathrm{out}}} q_{\mathrm{out}}^{k}(t)},
\end{equation}
where $j$ and $k$ represent the counters for the elements of the set $L_{\mathrm{in}}$ of links flowing into the junction and a set $L_{\mathrm{out}}$ of links withdrawing flow from the junction, respectively; $q_{\mathrm{in}}^{j}(t)$ and $q_{\mathrm{out}}^{k}(t)$ are the corresponding inflows and outflows from these links; $c_\mathrm{in}^j(t)$ is the chemical concentration in each of the inflows;  $q^\mathrm{B_\mathrm{J}}_i(t)$ is the flow injected into the junction with chemical concentration $c^\mathrm{B_\mathrm{J}}_i(t)$ by a booster station, if located; and $q^{\mathrm{D}_\mathrm{J}}_i(t)$ represents the consumer's demand.

\subsubsection{Mass Balance at Tanks} In our model, we assume complete and instantaneous mixing of all inflows, outflows, and stored water in a tank, following the continuously stirred tank reactor (CSTR) model. Consequently,  

\begin{equation}\label{equ:tank2}
	\begin{split}
	 & V_i^\mathrm{TK}(t + \Delta t) c_i^\mathrm{TK}(t+ \Delta t) = V_i^\mathrm{TK}(t) c_i^\mathrm{TK}(t) + \\ & \sum_{j \in L_{\mathrm{in}}} q^j_\mathrm{in}(t)c^j_\mathrm{in}(t) \Delta t  +  V^\mathrm{B_\mathrm{TK}}_i(t+\Delta t)c^\mathrm{B_\mathrm{TK}}_i(t+\Delta t) - \\
		& \sum_{k \in L_{\mathrm{out}}} q^k_\mathrm{out}(t)c_i^\mathrm{TK}(t) \Delta t +R^\mathrm{TK}_{\mathrm{MS}}(c_i^\mathrm{TK}(t)) V_i^\mathrm{TK}(t) \Delta t,
	\end{split}
\end{equation}
where $V_i^\mathrm{TK}(t)$ is the tank volume and $V^\mathrm{B_\mathrm{TK}}_i(t+\Delta t)$ is the chemical solute volume injected to the tank by a booster station, if located, with a concentration $c^\mathrm{B_\mathrm{TK}}_i(t+\Delta t)$. Note that the effect of booster station injections is considered immediate according to the CSTR model. Therefore, their volume and concentration are accounted for at $t + \Delta t$ to calculate the tank volume concentration at the same time instant. $R^\mathrm{TK}_{\mathrm{MS}}(c^\mathrm{TK}_i(t))$ is the multi-species dynamics in tanks expression (refer to Section \ref{sec:MSmodel}).

\subsection{Chlorine Multi-Species Reaction and Decay Dynamics and DBPs Formation Models}~\label{sec:MSmodel} 
In this paper, we utilize a first-order decay dynamics model for chlorine. Additionally, we model the mutual reaction between chlorine and a fictitious reactant using a second-order formulation, which also considers the formation of DBPs. Specifically, our focus is on the formation of one of the most common types of DBPs: trihalomethanes (THMs) \cite{DisinfectionByproductsDBPs2022}. The chlorine decay reaction rates for Pipe $i$ and Tank $j$ are $k_i^\mathrm{P} = k_{b}+\frac{2k_{w}k_{f}}{r_{\mathrm{P}_i}(k_{w}+k_{f})},\,\,\,\,\, k_j^\mathrm{TK} = k_{b}$, where $ k_{b}$ is the bulk reaction rate constant; $k_{w}$ is the wall reaction rate constant; $k_{f}$ is the mass transfer coefficient between the bulk flow and the pipe wall; $r_{\mathrm{P}_i}$ is the pipe radius. Both decay rates are in 1/sec. 


By donating the chlorine concentration to be $c$, \textcolor{cadmiumgreen}{$\tilde{c}$} for fictitious reactant concentrations, and  \textcolor{persimmon}{$\hat{c}$} for THMs, the multi-species dynamics models \cite{moeiniBayesianOptimizationBooster2023} in pipes and tanks are expressed as in Tab. \ref{tab:MSExp}. In this table, $k_r$ (L$\cdot$mg$^{-1}\cdot$sec$^{-1}$) denotes the mutual reaction coefficient, while $Y_\mathrm{FR}$ and $Y_\mathrm{THMs}$ represent the unitless ratios between the stoichiometric coefficients of the fictitious reactant and THMs, respectively, to that of chlorine.

\subsection{Chlorine and Byproducts Multi-Species Dynamics in a Form of State-Space Representation}

The state-space representation of the WQ multi-species dynamics of chlorine and byproducts (WQMS-CLBP) is expressed in Eq. \eqref{eq:WQMSCLBP-NLDE} as nonlinear difference equations (NLDE).  

\begin{Rep}
	\noindent\colorbox{darkcerulean}{\textcolor{white}{\textbf{WQMS-CLBP}}} \hfill {\raggedleft \colorbox{darkcerulean}{\textcolor{white}{\textbf{NLDE}}}\par} 
	\vspace{-4mm}
	\begin{subequations}~\label{eq:WQMSCLBP-NLDE}
		\begin{align}
					 \mE(t) \vx(t+\Delta t) &= \mA(t) \vx(t) + \mB(t) \vu(t) + \vf(\vx(t)), \\ 
					 \vy(t) &= \mC(t) \vx(t),
		\end{align}
	\end{subequations}
\end{Rep}
where $\vx(t)$ represents the state vector, which concatenates the concentrations of chlorine, fictitious reactant, and the byproducts at the various components within the network. The control input vector, $\vu(t)$, encompasses chlorine injections and can also accommodate unplanned and planned injections of contamination associated with the fictitious reactant. The nonlinear vector, $\vf(\vx(t))$, encapsulates the expressions for mutual reactions and byproduct formation. Vector $\vy(t)$ contains sensor measurements of the chemicals. Matrices $\mE(t), \mA(t), \mB(t),$ and $\mC(t)$ are time-varying and depend on factors such as network topology, hydraulic parameters, decay rates, coefficients for mutual reactions between chemicals, and the locations of booster stations and sensors.

Note that, matrix $\mE(t)$ equals identity under the condition of applying explicit discretization schemes. However, for the implementation of implicit discretization schemes, its construction depends on the system's hydraulics (refer to Tab. \ref{tab:DiscSchms}). The system's hydraulics are updated every hydraulic time-step, which is typically longer than the WQ one. Therefore, it is customary to donate the time instant when this matrix is taken as $t$. 

As explained, the one source of nonlinearity in our model is the mutual reaction dynamics between the chemicals. To overcome the complexity associated with this nonlinearity, we employ a linearization technique, specifically utilizing Taylor series approximation as detailed in \cite{elsherifComprehensiveFrameworkControlling2024}. Herein, we showcase the linearization process for the mutual reaction expression, which applies uniformly across all such expressions listed in Tab. \ref{tab:MSExp}. The linearization is performed around operating points represented by ${c}_\mathrm{o}$ for chlorine and \textcolor{cadmiumgreen}{$\tilde{c}_\mathrm{o}$} for the fictitious reactant.

\begin{equation}
	\begin{split}
			& R_{\mathrm{MS}}(\textcolor{cadmiumgreen}{\tilde{c}(t)})  = - Y_\mathrm{FR} k_r c(t) \textcolor{cadmiumgreen}{\tilde{c}(t)} \\
			& \Rightarrow - Y_\mathrm{FR} k_r \Big(c_\mathrm{o} \textcolor{cadmiumgreen}{\tilde{c}(t)} + \textcolor{cadmiumgreen}{\tilde{c}_\mathrm{o}} c(t) - {c}_\mathrm{o} \textcolor{cadmiumgreen}{\tilde{c}_\mathrm{o}} \Big).
	\end{split}
\end{equation}

For both chlorine and the fictitious reactant, the linearized mutual reaction breakdown consists of terms dependent on their respective concentrations, terms dependent on the concentrations of the other chemical, and constant terms. While THMs concentrations depend on the other two chemicals concentrations and a constant term. Consequently, while the general state-space representation in Eq. \eqref{eq:WQMSCLBP-NLDE} features a block-diagonal matrix of $\mA$ matrices with no interdependency between the chemicals apart from within the $\vf$ function, the application of linearization alters the state-space representation to linear difference equations (LDEs) with interdependencies among these chemical compounds. This transformation can be expressed as:

\begin{Rep}
	\noindent\colorbox{darkcerulean}{\textcolor{white}{\textbf{WQMS-CLBP}}} \hfill {\raggedleft \colorbox{darkcerulean}{\textcolor{white}{\textbf{LDE}}}\par} 
	\vspace{-4mm}
	\begin{subequations}~\label{eq:WQMSCLBP-LDE}
		\begin{align}
			\mE(t) \vx(t+\Delta t) &= \tilde{\mA}(t) \vx(t) + \mB(t) \vu(t) + \boldsymbol{\Phi}, ~\label{eq:WQMSCLBP-LDEa} \\ 
			\vy(t) &= \mC(t) \vx(t),
		\end{align}
	\end{subequations}
\end{Rep}
where $\tilde{\mA}(t)$ is the updated matrix to account for the interdependencies among the chemicals after linearization, while $\boldsymbol{\Phi}$ is the vector gathering the constant terms. 

The operating points around which the system is linearized are dynamically updated throughout the simulation period, typically every other time-step. The updating process is guided by the frequency with which the system dynamics evolve. In practice, the operating points are updated on a time scale wider than the WQ time-step. This approach ensures that the linearization captures the evolving behavior of the system adequately, allowing for accurate modeling of the system dynamics over time.
	
\section{Disinfectant Control Problem Formulation}~\label{sec:CLDBPCntrlProbForm}

The primary objective of this study is to regulate disinfectant levels within WDNs to comply with standard thresholds, while simultaneously addressing the formation of DBPs. Our proposed approach involves the implementation of a MPC algorithm, which is constructed based on the WQ multi-species model outlined in \eqref{eq:WQMSCLBP-LDE}, and constrained by specified chlorine limits and DBPs cutoffs. As a preliminary step, we conduct a controllability analysis of the system to identify critical states (e.g., dead-ends) and determine control inputs from booster stations. Utilizing controllability metrics obtained from this analysis, we assign higher weights to control inputs that exhibit greater effectiveness in influencing critical states or achieving desired objectives. 


First, we explain the formulation of the disinfectant control problem. The control problem is formulated over the simulation period $[0,T_\mathrm{s}]$, with the objective of minimizing the cost of chlorine injections. The problem is constrained by multiple constraints, including maintaining the chlorine concentrations within the standard levels of 0.2 mg/L and 4 mg/L, and THMs levels lower than 0.08 mg/L. Additionally, the problem can be utilized to constrain the fictitious reactant concentrations to a specific level within the simulation period. The control inputs for chlorine are constrained to be non-negative and limited by the availability of chlorine and the capacity of booster stations. All these constraints must be satisfied while complying with the actual governing equations of the WQ dynamics. Combining all this information, the water quality multi-species control problem (WQMS-CP) is formulated as shown in \eqref{eq:opt}.

\begin{Rep}
	\noindent\colorbox{darkcerulean}{\textcolor{white}{\textbf{WQMS-CP}}} 
	\vspace{-5mm}
\begin{equation}~\label{eq:opt}
	\begin{aligned}
		\underset{\vx(t),\hat{\vu}(t)}{\mbox{minimize}} \hspace{1cm} & \mathcal{J}(\hat{\vu}(t)) = \epsilon \sum_{t=1}^{N_\mathrm{s}} \vq^\mathrm{B}(t)^\top \hat{\vu}(t) \\
		\mbox{subject to} \hspace{1cm} & \mbox{WQMS-CLBP} \; \eqref{eq:WQMSCLBP-LDE}, \\ & \vx_{\min} \leq \vx(t) \leq \vx_{\max}, \\ & \hat{\vu}_{\min} \leq \hat{\vu}(t) \leq \hat{\vu}_{\max},
	\end{aligned}		
\end{equation}
\end{Rep}
\vspace{-2mm}
where problem variables $\vx(t)$ and $\tilde{\vu}(t)$ are chemicals concentrations network-wide and chlorine injections through booster stations, $\vq^\mathrm{B}(t)$ is the flow rates at the nodes corresponding to the locations of the booster stations, $\epsilon$ is the unit cost of chlorine in \$/mg, and WQMS-CLBP is the WQ model we are simulating and controlling following the representation in \eqref{eq:WQMSCLBP-LDE}. Finally, $N_\mathrm{s}$ is the number of time-step in the simulation period, $N_\mathrm{s} = \frac{T_\mathrm{s}}{\Delta t}$.

The control problem described in \eqref{eq:opt} is formulated as a linear program (LP) due to its linear objective function and constraints. The subsequent step involves reformulating this problem into a quadratic program (QP) with a quadratic objective function in a MPC framework. This reformulation aims to minimize chlorine injections while ensuring the smoothness of control inputs while ensuring falling within the states boundaries. The study \cite{wangHowEffectiveModel2021b} has provided the comprehensive derivation of this problem for single-species WQ dynamics based on AR models. Despite focusing on a different system, the derivation is applicable to our study since both systems are linear, and the form of the objective functions and constraints remains consistent. The one distinction lies in the constant terms concatenated in $\boldsymbol{\Phi}$ of \eqref{eq:WQMSCLBP-LDEa}. However, these terms solely depend on the operating point around which the system is linearized. 
In accordance with the approach outlined in \cite{wangHowEffectiveModel2021b}, the full derivation and formulation of the control problem are provided therein. For brevity, we direct readers to this study for a detailed understanding of the derivation and the final formulation of the control problem.

In this final formulation of the control problem as presented in \cite[Eq. (34)]{wangHowEffectiveModel2021b}, two weight matrices are introduced: $\mQ=\mQ^\top$ and $\mR=\mR^\top$. These matrices specify the relative importance of measurement deviations and the smoothness of control inputs, respectively. In addition, we are building these matrices to reflect on prioritizing each of control input according to the controllability analysis we preform beforehand. Detailed information on the controllability analysis and the construction of these matrices is provided in the next section.

%

\subsection{Preliminary Controllability Analysis}

In this section, we present the procedure for conducting the prior controllability analysis and deriving the corresponding weight matrices of the MPC algorithm. To do so, we start by introducing the notion of controllability for dynamic systems, focusing specifically on the WQ dynamic system in our study. We then introduce metrics aimed at quantifying the influence of each booster station within the control framework. We note that the controllability notion introduced in this section is based on the linearized form of the WQMS-CLBP model as a simplified approximation, yet, enough to provide the needed insights for the analysis. 

From a control-theoretic perspective, controllability refers to the capability of guiding a system from its initial states $\m{x}_{o}:=\m{x}_{t(0)}$ to a desired state $\m{x}_{p}:=\m{x}_{T_\mathrm{p}}$ by some input $\m{u}(t)$ over a specific time window of $T_\mathrm{p}$~\cite{kalmanMathematicalDescriptionLinear1963}. In the context of WQ control, we want to measure the ability to regulate chlorine concentrations by adjusting chlorine injections from booster stations, ensuring chlorine levels remain within predefined bounds.

The dynamic linear system \eqref{eq:WQMSCLBP-LDE}, where $\tilde{\mA} \in \mathbb{R}^{n_x \times n_x}$ and $\mB \in \mathbb{R}^{n_x \times n_u}$, is said to be controllable if only if the controllability matrix for $N_\mathrm{p}= \frac{T_\mathrm{p}}{\Delta t}$ time-steps given as
\begin{equation}~\label{eq:control_matrix}
	\mathcal{C}_{N_\mathrm{p}} := \{
		\mB, \;\; \tilde{\mA} \mB, \;\; \tilde{\mA}^{2} \mB, 
		\ldots, \;\; \tilde{\mA}^{N_\mathrm{p}-1} \mB \}
	\in \mathbb{R}^{n_{x}\times N_\mathrm{p}n_{u}},
\end{equation}
is full row rank, i.e, $\mr{rank}(\mathcal{C}_{N_{p}}) = n_x$, without loss of generality as we assume that $N_\mathrm{p}n_{u} > n_{x}$. This is known as Kalman's rank condition~\cite{kalmanMathematicalDescriptionLinear1963}. 

For our analysis objective, simply assessing the rank property is not sufficient to evaluate the impact of each booster station's injections within the control framework. The concept of control energy is important as well. In the WQ control context, control energy relates to the amount of chlorine injections to reach the desired chlorine levels at the networks' components.  Metrics concerning the control energy are derived from the controllability Gramian $\m{W}_{c}(\tilde{\m{A}},\m{B},N_\mathrm{p}):= \m{W}_{c} \in \mathbb{R}^{n_{x}}$, defined for $N_\mathrm{p}$ pairs of matrices $\tilde{\m{A}}$ and $\m{B}$ as
\begin{equation}~\label{equ:control_gram}
	\hspace{-0.5cm}	\m{W}_{c} := \sum_{\tau=0}^{N_\mathrm{p}-1}\tilde{\m{A}}^{\tau}\m{B} \m{B}^{\top}(\tilde{\m{A}}^{\top})^{\tau} = 
	\mathcal{C}_{N_\mathrm{p}}\mathcal{C}_{N_\mathrm{p}}^{\top},
\end{equation}
where the controllability Gramian $\m{W}_{c}$, that is a positive semidefinite matrix. In this study, we employ the $\mathrm{trace}(\mW_{c})$ metric, which inversely correlates with the average controllability energy across all state-space directions. In the context of WQ control, a higher
controllability energy indicated a greater potential for chlorine injections to impact various system states within the specified time interval. In addition, rather than utilizing the $\mathrm{rank}$ metric for the controllability matrix in higher dimensions, we apply it for the controllability Gramian--a symmetric matrix with lower dimensions. 

However, there are two aspects demand attention during our controllability analysis: \textit{(i)} not all the states of the system dynamics \eqref{eq:WQMSCLBP-LDE} are critical or controllable by booster stations, and \textit{(ii)} calculating the $\mathrm{trace}$ for an uncontrollable subspace can be misleading due to the averaging of energy across uncontrollable directions. The first is addressed by adopting the concept of \textit{target controllability} \cite{gokhaleOptimizingControllabilityMetrics2021}. 
 Target controllability enables us to specify the desired target nodes, thereby mitigating the challenge of high dimensionality associated with the WQ representation. In this scenario, the metrics are applied to the {targeted controllability Gramian} $\boldsymbol{W}_{\mathcal{T}}=\mC_\mathcal{T} \m{W}_c \mC_\mathcal{T}^\top$, where the output matrix $\mC_\mathcal{T}$ identifies the set of
 target nodes $\mathcal{T}$ of size $n_t$. The notions of controllability and control energy apply to the subspace of the target nodes. Regarding the second aspect, to quantify the control energy for the controllable subspace within the space under consideration---where the rank of the controllability Gramian is $k < n$, with $n$ denoting the number of states in the original space---a decomposition approach is utilized \cite{dattaNumericalMethodsLinear2004}. This approach begins by defining a nonsingular matrix $\mT \in \mathbb{R}^{n \times n}$ such that  
\begin{equation}~\label{eq:UncontrlSysDecomp}
	\begin{split}
		 \bar{\mA} =  \mT \tilde{\mA} \mT^{-1} = \begin{bmatrix}
			\bar{\mA}_{11} & \bar{\mA}_{12} \\
			\boldsymbol{0} & \bar{\mA}_{22}
		\end{bmatrix}, \;\; \bar{\mB} = \mT \mB = \begin{bmatrix}
			{\bar{\mB}_{1}} \\ \boldsymbol{0}
		\end{bmatrix},
	\end{split}
\end{equation}
where $\bar{\mA}_{11}, \; \bar{\mA}_{12}$ and $\bar{\mA}_{22}$ have dimensions of $k \times k, \; k \times (n-k)$ and $(n-k) \times (n-k)$, and $\bar{\mB}_1$ has $k$ rows. Matrices $\bar{\mA}_{11}$ and $\bar{\mB}_{1}$ define a controllable subspace.

The key question to address now is: \textit{over which time interval should the WQ controllability metrics be evaluated, and correspondingly, how often should the MPC weight matrices be updated?} To answer this question, we emphasize that this time interval should coincide with the frequency at which booster station injections' effectiveness varies. This effectiveness is dependent on the changes in system dynamics, particularly system's hydraulics. Hence, the time interval is taken to align with the hydraulic time-step, during which the WQ time-step is updated due to their different scales.

{	\begin{algorithm}[h!]
		\small	\DontPrintSemicolon
		\SetAlgorithmName{Procedure}
		1 1\KwIn{WDN topology, components’ characteristics, hydraulics
			parameters, the linearized WQMS-CLBP system matrix $\tilde{\mA}$, and WQ initial conditions}
		\KwOut{MPC weight matrices $\mR(t)$ and $\mQ(t)$ at each hydraulic time-step} 
		\Init{}{ Define critical $n_\mathcal{T}$ sets of target nodes $\mathcal{T}_i, i=1, \ldots, n_\mathcal{T}$  \;
			Assign index $\eta_i, i=1, \ldots, n_\mathcal{T}$ to each target set based on their criticality/importance \;
			Obtain $\Delta t,$ $s_{\mathrm{L}_i}$ and $\Delta x_i,$ $\forall \; i=1, \ldots, n_\mathrm{P}$ via following Procedure \ref{alg:Proc1} \;
		}
		\For{each hydraulic time-step}{
			\For{$j= 1$ \textbf{to} ${n_{\hat{u}}}$}{
				Initialize $w_j \leftarrow 0$ \;
				Construct $\mB_j$ asumming only booster station $j$ is allocated  \textcolor{blue}{// $n_{\hat{u}}$ is booster stations count} \;
				Construct $\mathcal{C}_{N_\mathrm{p}}$ via \eqref{eq:control_matrix} \; 
				Construct $\m{W}_{c}$ via \eqref{equ:control_gram} \; 
				\For{$k= 1$ \textbf{to} ${n_{\mathcal{T}}}$}{
					Calculate $\boldsymbol{W}_{\mathcal{T}_k}=\mC_{\mathcal{T}_k} \m{W}_c \mC_{\mathcal{T}_k}^\top$ \;
					Calculate $\mathrm{rank}(\boldsymbol{W}_{\mathcal{T}_k})_j$\;
					\uIf{$\mathrm{rank}(\boldsymbol{W}_{\mathcal{T}_k})_j = n_{t_k}$}{
						Calculate $T_r=\mathrm{trace}(\boldsymbol{W}_{\mathcal{T}_k})_j$ \;
						Update $w_j \leftarrow w_j + \eta_k \cdot n_{t_k} \cdot T_r$ \;
					}\Else{$r = \mathrm{rank}(\boldsymbol{W}_{\mathcal{T}_k})_j$ \;
						Apply subspace decompositon as in \eqref{eq:UncontrlSysDecomp} \;
						Update $\widetilde{\mC}_{\mathcal{T}_k}$ to the first $r$ of $ \mT \mC_{\mathcal{T}_k}  \mT^{-1}$\;
						$\widetilde{\boldsymbol{W}}_{{\mathcal{T}_k}_j} \leftarrow \widetilde{\mC}_{\mathcal{T}_k} \m{W}_c \widetilde{\mC}_{\mathcal{T}_k}^\top$  \;
						Calculate $T_r=\mathrm{trace}(\widetilde{\boldsymbol{W}}_{{\mathcal{T}_k}_j})$ \; 
						Update $w_j \leftarrow w_j + \eta_k \cdot r \cdot T_r$
					}
				}
			}
			Calculate diagonal elements $a_{jj}, j = 1, \ldots, n_{\hat{u}}$ of $\mR(t)$, $a_{jj} = w_j/ \sum_{j=1}^{n_{\hat{u}}}w_j$\;
			Assign diagonal weights to $\mQ(t)$ \textcolor{blue}{// According to critical states and importance in comparison to $\mR(t)$} 
		}			
		\caption{Offline prior controllability analysis and MPC weight matrices determination\label{alg:Proc2}}
\end{algorithm}}

After establishing the controllability metrics to be utilized, herein, we outline the analysis conducted to determine the weights for the MPC matrices. Initially, for each hydraulic time-step, the matrix $\mB$ is constructed for each booster station individually as the sole station in the system, followed by the computation of the corresponding controllability matrix and Gramian. Subsequently, for each critical set of target nodes $\mathcal{T}$ and each booster station, the $\mathrm{rank}$ and $\mathrm{trace}$ metrics of the Gramians are calculated, leveraging the proposed methodologies addressing target controllability and uncontrollable subspace decomposition. Next, based on the importance index assigned to each critical set, each booster station is allocated a score relative to others according to its WQ controllability metrics. These scores are then aggregated for each booster station and weighted together relatively. Thus, the weights to construct the matrix $\mR(t)$ are computed for each hydraulic time-step.

On the other hand, matrix $\mQ$ is constructed relative to matrix $\mR$. Essentially, if prioritizing the smoothness of control inputs and the distinction between booster station injections is of utmost importance, the weights in $\mR$ are proportionally higher. Conversely, if achieving desired chemical levels promptly is more important, the weights in $\mQ$ are elevated. Such decisions are guided by the deviation of current concentrations from the desired setpoints and the how effective are the booster stations injections to cover the network. Typically, tuning these matrices is performed individually for each network to achieve suitable settings.

All these steps are summarized in Procedure \ref{alg:Proc2}. Note that, this approach is flexible, allowing for consideration of either one metric or both in the assessment and weighting procedure, albeit with different weights. Additionally, the critical sets of target nodes are categorized to be one of the following: dead-ends, zones with low initial chlorine conditions compared to the rest of the network, zones with high contaminant concentrations, or zones with pre-existing elevated levels of byproducts. For the latter scenario, the assigned index to this set is the lowest.

%

%
%
%

\begin{figure*}[t!]
	\centering
	\includegraphics[width=\textwidth]{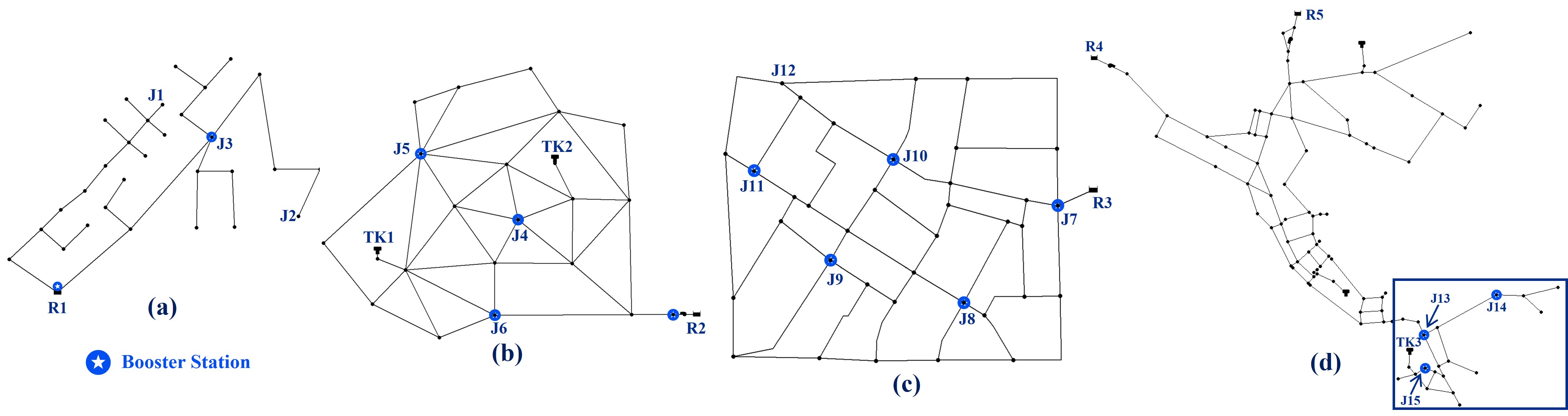}
	\vspace{-0.5cm}
	\caption{ Networks under study and their layouts: (a) BLA-M, (b) Anytown, (c) FOS, and (d) Net3 networks.}~\label{fig:CaseStudy}
\end{figure*}

\section{Case Studies}~\label{sec:CaseStudies}
In this section, we validate the proposed chlorine control approach through numerical case studies. These studies include several networks characterized by different scales, layouts, hydraulic settings and scenarios, and initial WQ conditions. The hydraulic settings are obtained by running different scenarios using the EPANET toolkit on MATLAB \cite{rossmanEPANETUserManual2020}. We apply our approach on four benchmark networks, each has a different number of components and layouts, including looped networks and those with a relatively higher number of dead-ends. These networks are: a modified version of the Blacksburg network (BLA-M), Anytown, Fossolo network (FOS), and Net3 \cite{wangTwoObjectiveDesignBenchmark2015,rossmanEPANETUserManual2020}. Fig. \ref{fig:CaseStudy} illustrates these networks layouts while Tab. \ref{tab:NetworksComponents} lists the count for each component. 

\begin{table}[h!]
	\centering
	\begin{threeparttable}
	\caption{Components count for each of the test networks.}~\label{tab:NetworksComponents}
	\begin{tabular}{c|c|c|c|c|c}
		\hline
	Network & Junctions & Reservoirs & Tanks & Pipes & Pumps \\
		\hline
		BLA-M & 30 & 1 & 0 & 30 & 0 \\
		Anytown & 22 & 1 & 2 & 43 & 3 \\
		FOS & 36 & 1 & 0 & 58 & 0 \\
		Net3 & 90 & 2 & 3 & 114 & 2 \\
		\hline
		\hline
	\end{tabular}
\end{threeparttable}
\end{table}

\begin{figure}[h!]
	\centering	\subfloat[\label{fig:BLA_J1}]{\includegraphics[keepaspectratio=true,width=0.35\textwidth]{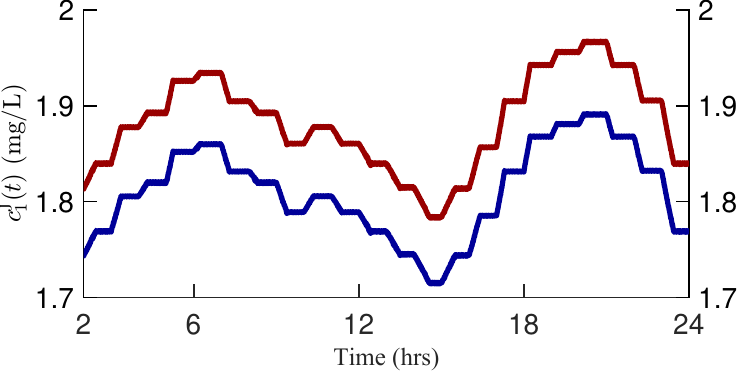}}{}
	\subfloat[\label{fig:BLA_J2}]{\includegraphics[keepaspectratio=true,width=0.35\textwidth]{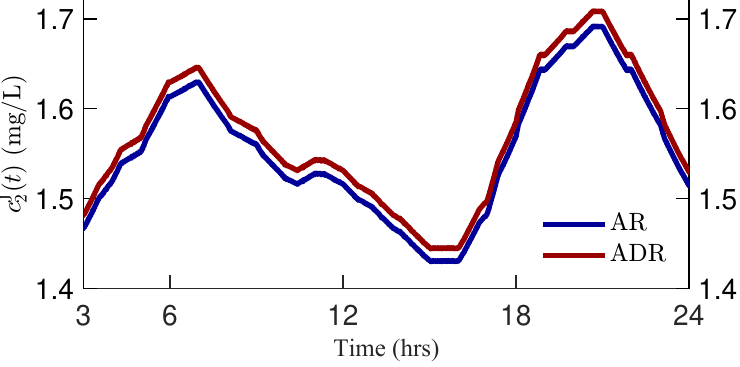}}{}
	\caption{Chlorine concentrations at Junctions (a) J1 and (b) J2 of BLA-M network without (AR) and with (ADR) the consideration of the dispersion process effect.}~\label{fig:BLA_Disp}
	\vspace{-0.3cm}
\end{figure}

First, we demonstrate the effect of considering dispersion in the simulation of the chemicals transport and evolution. We simulate the three chemical compounds with and without the inclusion of dispersion on the BLA-M and Anytown networks for comparison. For both networks, the simulation period is 24 hours with a hydraulic time-step of 1 hour. In addition, fixed sources of chemicals are maintained at Reservoirs R1 and R2, with concentrations of 2 mg/L for chlorine, 0.3 mg/L for the fictitious reactant, and 0.01 mg/L for the THMs. For the BLA-M network, Fig. \ref{fig:BLA_Disp} shows the results for chlorine concentrations at Junctions J1 and J2 with and without dispersion. As shown in the figure, the inclusion of dispersion has a greater impact on Junction J1 than J2. Neglecting dispersion leads to an underestimation of chlorine concentrations, particularly at J1, with an underestimation of around 8\%, while for J2, it is 3\%. This is due to the lower velocity in the pipe leading to Junction J1 compared to J2, resulting in a more dispersion-dominant process. 

\begin{figure}[h!]
	\centering	\subfloat[\label{fig:Anytown_TK1Cl}]{\includegraphics[keepaspectratio=true,width=0.4\textwidth]{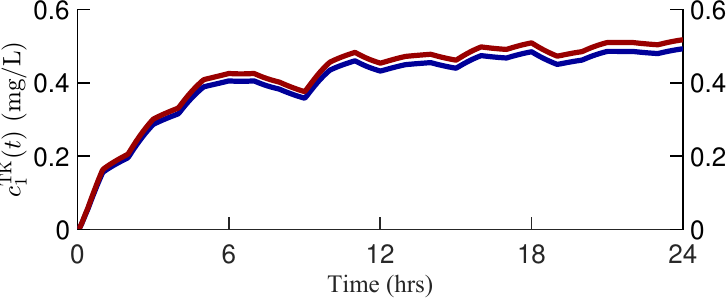}}{}\hspace{0.3cm}
	\subfloat[\label{fig:Anytown_TK1THMs}]{\includegraphics[keepaspectratio=true,width=0.4\textwidth]{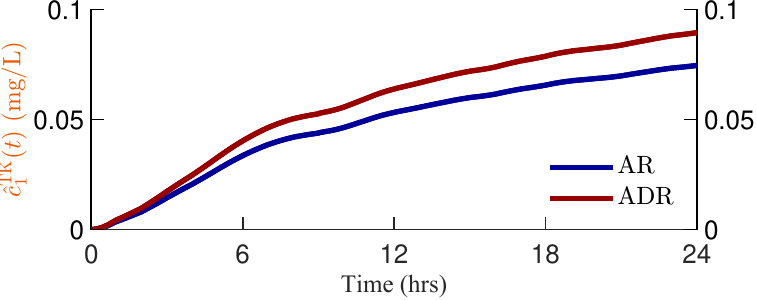}}{}
	\caption{(a) Chlorine and (b) THMs concentrations at Tank TK1 of the Anytown network without (AR) and with (ADR) the consideration of the dispersion process effect.}~\label{fig:Anytown_Disp}
\end{figure}

For the Anytown network, Fig. \ref{fig:Anytown_Disp} illustrates chlorine and THMs concentrations at Tank TK1 with and without dispersion. The reason for showing these results is to highlight that although the underestimation of chlorine concentrations at Tank TK1 may seem negligible, the difference in THMs concentrations can be substantial, potentially leading to overlooking reaching its higher bound, thus compromising water safety. This situation arises from how mutual reactions are handled in our model. These reactions are expressed as second-order nonlinear formulations. Consequently, the dispersion effect on chlorine and the fictitious reactant implicitly affects THMs formation while explicitly impacting its evolution in the transport expression. This effect becomes more noticeable with higher concentrations of chlorine and the fictitious reactant, as well as with slower velocities.

It is worth mentioning that the results in Fig. \ref{fig:BLA_Disp} are obtained by applying the explicit discretization schemes listed in Tab. \ref{tab:DiscSchms}, while the results in Fig. \ref{fig:Anytown_Disp} are obtained using the implicit discretization schemes. Additionally, Procedure \ref{alg:Proc1} is employed to determine the WQ time-step and system dimensions for both networks under different scenarios to test its applicability and numerical stability. Although the discretization procedure has proven its applicability, in some scenarios, the resulting WQ time-step is restrictively small to satisfy the condition listed in step 14 of Procedure \ref{alg:Proc1}. Subsequently, applying explicit discretization schemes leads to a high number of segments required for each pipe, resulting in high system dimensions. This high system dimensionality demands significant computational time for the simulation of chemical evolution, considering that we simulate for three chemicals. Thus, for each network component and pipe segment, we have three states. Conversely, using an implicit discretization scheme for either the AR-PDE or ADR-PDE avoids this issue but requires performing matrix inversion. This problem can be mitigated by constructing these matrices as sparse matrices and utilizing fast inversion commands according to the coding language used. To that end, the results obtained in the remainder of this section are based on models developed using implicit discretization schemes.

\begin{figure}[h!]
	\centering	\subfloat[\label{fig:FOS_Cont1}]{\includegraphics[keepaspectratio=true,width=0.4\textwidth]{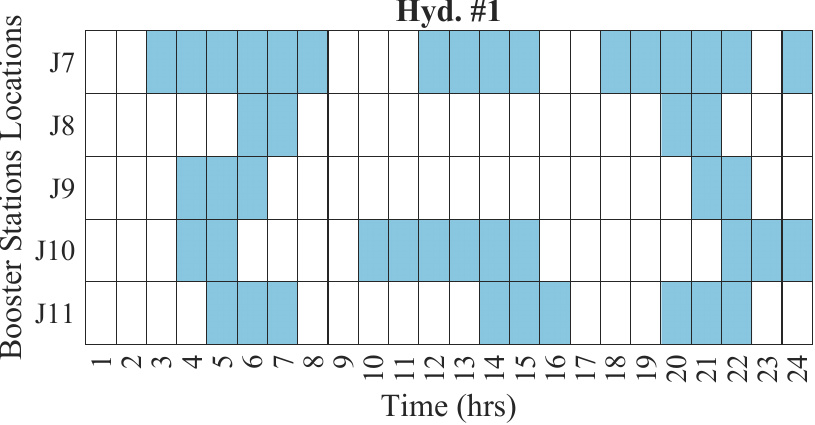}}{}\hspace{0.4cm}
	\subfloat[\label{fig:FOS_Cont2}]{\includegraphics[keepaspectratio=true,width=0.4\textwidth]{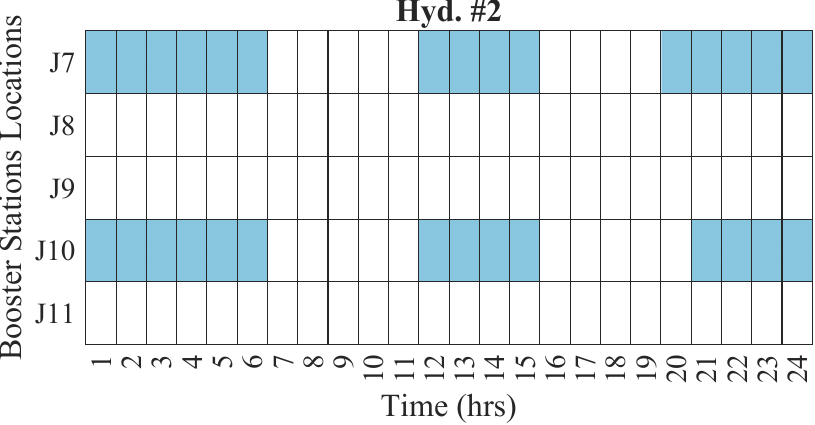}}{}
	\caption{Controllability of chlorine injections by each booster stations allocated over the FOS network if worked solely to steer the concentrations at the target junction J12, for two hydraulic settings: (a) Hyd. \#1 and (b) Hyd. \#2. Colored tiles indicate full $\mathrm{rank}$ metric. \vspace{-0.2cm}}~\label{fig:FOS_Cont}
\end{figure}

Next, we present results from quantifying the controllability of each booster station located on the FOS network individually, aimed at steering chlorine concentration to a critical target node: Junction J12. This junction could represent a high-demand area or an area with initially low chlorine concentrations. We calculate the $\mathrm{rank}$ metric for each booster station and present the results in Fig. \ref{fig:FOS_Cont} for two hydraulic settings of the system over a 24-hour simulation period with a hydraulic time-step of 1 hour. The WQ time-step is updated within the hydraulic time-step on a scale of 1 minute. These metrics are measured on a system linearized around operating points of 0.5 mg/L for chlorine, 0.1 mg/L for the fictitious reactant, and 0.01 mg/L for THMs concentrations---the initial WQ concentrations of the system. 

In Fig. \ref{fig:FOS_Cont}, achieving full $\mathrm{rank}$ is highlighted by the colored blocks within the specific hydraulic 1-hour time-step,  demonstrating the achieved target controllability for Junction J12. As shown, Junction J12 is not always controllable by booster stations due to changes in flow directions and actual flow rates, which can make it difficult to reach the junction within the specified time interval. For the first hydraulics scenario (Hyd. \#1), Junction J12 can be influenced by different booster stations at various time-steps, with varying control energy depending on the flow paths between them. Conversely, for the second hydraulics scenario (Hyd. \#2), higher demands exist on the opposite side of the network, causing flow directions to be opposite to those leading to J12, with low flow rates to J12 except during the controllable windows shown in Fig. \ref{fig:FOS_Cont2}. This suggests that the other side of the network might also represent critical target nodes that demand higher priority indices.

\begin{figure}[h!]
	\centering
	\subfloat[\label{fig:BLAM_MPCU}]{\includegraphics[keepaspectratio=true,width=0.35\textwidth]{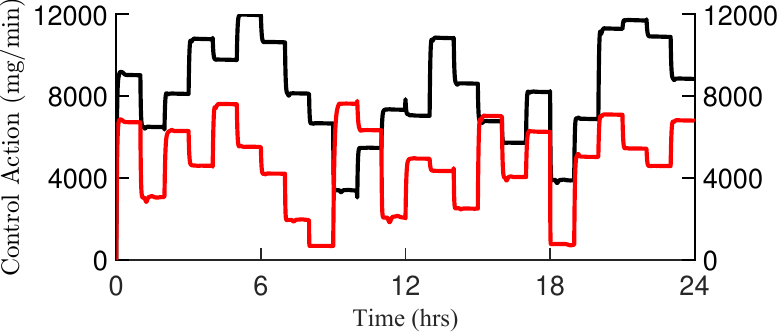}}{}	\subfloat[\label{fig:BLAM_MPCJ1}]{\includegraphics[keepaspectratio=true,width=0.35\textwidth]{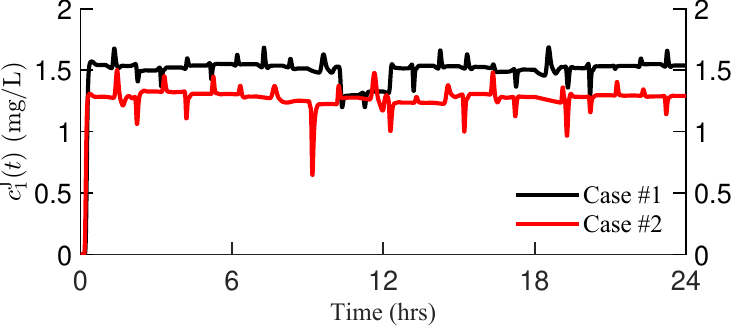}}{}
	\subfloat[\label{fig:BLAM_MPCJ2}]{\includegraphics[keepaspectratio=true,width=0.35\textwidth]{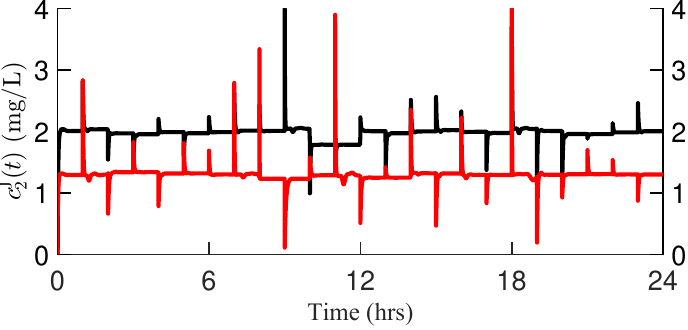}}{}
	\caption{(a) Summation of MPC control actions at Reservoir R1 and Junction J3 of the BLA-M network, and the corresponding chlorine concentrations at (b) Junction J1 and (c) Junction J2 for two cases: Case \#1, applying MPC without prior controllability analysis and without DBPs constraints; and Case \#2, applying MPC with prior controllability analysis and DBPs constraints. }~\label{fig:BLAM_MPC}
\end{figure}

By applying the proposed MPC control approach based on the results from the controllability analysis and constrained by THMs formation (Case \#2), the results for the BLA-M network are demonstrated in Fig. \ref{fig:BLAM_MPC}. Additionally, this figure shows results for Case \#1, where the MPC control algorithm is applied without prior controllability analysis and without constraining the problem with maximum THMs concentrations. It is important to note that the results for Case \#2 are obtained by tuning the matrices $\mR$ and $\mQ$ to balance control input smoothness, achieve the desired concentrations, and weight the inputs appropriately. As illustrated in Fig. \ref{fig:BLAM_MPCU}, the combined control actions from R1 and J3 are lower for the second case, while maintaining chlorine concentrations between the lower and upper bounds across the network—see Fig. \ref{fig:BLAM_MPCJ1} and Fig. \ref{fig:BLAM_MPCJ2} for examples. Two significant factors explain these results. The first factor is the values and scaling of $\mR$ and $\mQ$ compared to the cost of chlorine injections, allowing the latter to dominate the objective functions with higher trade-offs for the first two. The second factor is the constraint on chlorine injections to prevent THMs formation from exceeding its maximum bound of 0.08 mg/L. Furthermore, in Case \#1, the control actions are primarily assigned to the booster station at R1, while in Case \#2, they are distributed between the stations at R1 and J3. This distribution is due to the higher indices assigned to the dead-ends downstream of J3, where initial chlorine concentrations are zero and the fictitious reactant has relatively high initial concentrations of 0.5 mg/L, representing a contamination intrusion event in this zone. Additionally, fluctuations in chlorine concentrations at J2 shown in Fig. \ref{fig:BLAM_MPCJ2} are due to the need to recover chlorine levels after their consumption by the fictitious reactant and withdrawn the water with high demands.

For the Net3 case study, we focus on the zone framed in Fig. \ref{fig:CaseStudy}. In this network, Tank TK3 faces periods of filling and periods of serving parts of the network throughout the simulation period. Thus, it is considered a critical target node during the filling windows to ensure the stored water has sufficient chlorine concentrations for safe distribution. As shown in Fig. \ref{fig:CaseStudy}, the nearest booster stations to this tank are located at Junctions J13 and J15. Notably, water reaching J15 must first pass through J13, which serves more dead-ends than J15. To accommodate these scenarios, a higher priority index is assigned to the target set leading to the tank during its filling windows. When the tank is emptying, it is considered uncontrollable, and its weight is set to zero in Procedure \ref{alg:Proc2}. This approach demonstrates the scalability and flexibility of the proposed control strategy, allowing it to be easily customized for each network and each unique scenario or special consideration that arises. 
\begin{figure}[h!]
	\centering
	\includegraphics[width=0.4\textwidth]{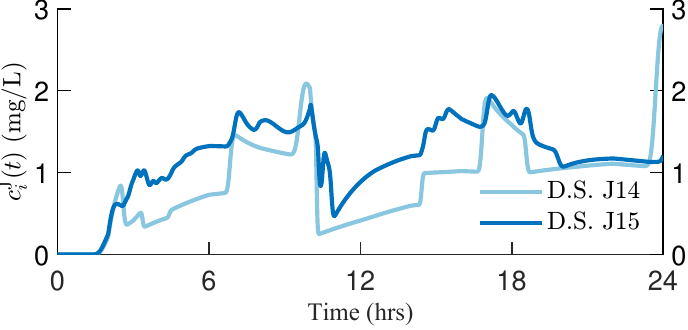}
	\caption{ Chlorine concentrations at the dead-ends downstream (D.S.) Junctions J14 and J15 after applying the MPC algorithm for Net3.}~\label{fig:Net3MPC}
\end{figure}

By applying the proposed MPC approach based on the aforementioned scenarios for Net3, Fig. \ref{fig:Net3MPC} illustrates chlorine concentrations at the dead-ends downstream Junctions J14 and J15. Results show how the water coming from Junction J13 with the same chlorine concentration but due to the different weights and interactive dynamics, the chlorine injections by booster stations at J14 and J15 are different. We also want to highlight that although Net3 is a relatively large scale network, the switching between the two discretization schemes for the cases of with and without dispersion has been smooth and the computational time for a simulation period of 24 hours has only increased by 3\% to run through the if condition that decide on the scheme used.


\section{Conclusion, Paper’s Limitations, and Recommendations for Future Work}~\label{sec:ConcLimRec}

This study introduces a novel disinfectant control approach that ensures the WQ safety through chlorine injections while mitigating the hazardous formation of its DPBs. This approach integrates an advanced multi-species WQ model that is based on the ADR dynamics, dynamically switching between discretization schemes to accurately simulate chemical transport according to the dominant process.  Accounting for the dispersion process in chemical evolution simulation can significantly impact results, particularly in low-velocity zones such as branched networks with dead-ends and stagnant water. Additionally, this study demonstrates the utilization of prior controllability analysis, enabling the storage of offline weight matrices that are updated every hydraulic time-step, reflecting the subsequent changes in the WQ controllability metrics. These matrices inform the MPC algorithm for the chlorine control problem, strategically prioritizing booster stations based on their influence on critical states and desired objectives. Consequently, optimal chlorine injections are allocated among booster stations based on the insights from the controllability analysis. The proposed control approach is validated on different numerical case studies with various scales, layouts, and hydraulic conditions.

In our study, we exclusively focus on chlorine as the disinfectant and accordingly refer to its dynamics as the WQ dynamics, given that chlorine is one of the most widely used disinfectants. However, there are other disinfection methods that differ in their composition, effectiveness, dynamics in WDNs, byproducts, and usage limitations. For a detailed comparison of these disinfection methods and their broader impacts, we refer readers to the study in \cite{tsitsifliDisinfectionImpactsDrinking2018}. Furthermore, our study acknowledges several limitations within its structure, including the high dimensionality associated with the developed approach, potentially leading to demanding computational times. However, model order reduction approaches are available to be utilized and integrated into the control framework as presented in \cite{elsherifComprehensiveFrameworkControlling2024}. Other limitations are the predetermined locations of booster stations and the prerequisite of detailed hydraulic settings before WQ control approach implementation. To address these limitations, future research direction for our group is to explore incorporating uncertainty in the hydraulic settings, modeling transient conditions to better reflect real-world scenarios, and adopting an integrated approach that considers optimizing both hydraulic and WQ dynamics. Additionally, it is worth noting that our controllability analysis is performed for the linearized system, representing an approximation of the actual system dynamics. 

\bibliographystyle{IEEEtran}
\bibliography{WQDBPs}

\begin{thebibliography}{10}
\providecommand{\url}[1]{#1}
\csname url@samestyle\endcsname
\providecommand{\newblock}{\relax}
\providecommand{\bibinfo}[2]{#2}
\providecommand{\BIBentrySTDinterwordspacing}{\spaceskip=0pt\relax}
\providecommand{\BIBentryALTinterwordstretchfactor}{4}
\providecommand{\BIBentryALTinterwordspacing}{\spaceskip=\fontdimen2\font plus
\BIBentryALTinterwordstretchfactor\fontdimen3\font minus
  \fontdimen4\font\relax}
\providecommand{\BIBforeignlanguage}[2]{{%
\expandafter\ifx\csname l@#1\endcsname\relax
\typeout{** WARNING: IEEEtran.bst: No hyphenation pattern has been}%
\typeout{** loaded for the language `#1'. Using the pattern for}%
\typeout{** the default language instead.}%
\else
\language=\csname l@#1\endcsname
\fi
#2}}
\providecommand{\BIBdecl}{\relax}
\BIBdecl

\bibitem{constableCenturyInnovationTwenty2003}
G.~Constable and B.~Somerville, \emph{A {{Century}} of {{Innovation}}: {{Twenty
  Engineering Achievements}} That {{Transformed Our Lives}}}.\hskip 1em plus
  0.5em minus 0.4em\relax Joseph Henry Press, Jan. 2003.

\bibitem{graymanwaltermHistoryWaterQuality2018}
W.~M. Grayman, ``History of {{Water Quality Modeling}} in {{Distribution
  Systems}},'' \emph{WDSA / CCWI Joint Conference Proceedings}, vol.~1, Jul.
  2018.

\bibitem{wangDisinfectionByproductFormation2013}
J.-J. Wang, X.~Liu, T.~W. Ng, J.-W. Xiao, A.~T. Chow, and P.~K. Wong,
  ``Disinfection byproduct formation from chlorination of pure bacterial cells
  and pipeline biofilms,'' \emph{Water Research}, vol.~47, no.~8, pp.
  2701--2709, May 2013.

\bibitem{zhouStabilityDrinkingWater2023}
Q.~Zhou, Z.~Bian, D.~Yang, and L.~Fu, ``Stability of {{Drinking Water
  Distribution Systems}} and {{Control}} of {{Disinfection By-Products}},''
  \emph{Toxics}, vol.~11, no.~7, p. 606, Jul. 2023.

\bibitem{shah2024recent}
M.~Shah, M.~A. Mazhar, S.~Ahmed, B.~Lew, and N.~Khalil, ``Recent trends in
  controlling the disinfection by-products before their formation in drinking
  water: A review,'' \emph{Drinking Water Disinfection By-products: Sources,
  Fate and Remediation}, pp. 177--192, 2024.

\bibitem{DisinfectionByproductsDBPs2022}
``Disinfection {{By-products}} ({{DBPs}}) {{Factsheet}} {\textbar} {{National
  Biomonitoring Program}} {\textbar} {{CDC}},''
  https://www.cdc.gov/biomonitoring/THM-DBP\_FactSheet.html, Jul. 2022.

\bibitem{kalita2024assessing}
I.~Kalita, A.~Kamilaris, P.~Havinga, and I.~Reva, ``Assessing the health impact
  of disinfection byproducts in drinking water,'' \emph{ACS Es\&t Water},
  vol.~4, no.~4, pp. 1564--1578, 2024.

\bibitem{abokifaWaterQualityModeling2016}
A.~A. Abokifa, Y.~J. Yang, C.~S. Lo, and P.~Biswas, ``Water quality modeling in
  the dead end sections of drinking water distribution networks,'' \emph{Water
  Research}, vol.~89, pp. 107--117, Feb. 2016.

\bibitem{jonkergouwVariableRateCoefficient2009}
P.~M.~R. Jonkergouw, S.-T. Khu, D.~A. Savic, D.~Zhong, X.~Q. Hou, and H.-B.
  Zhao, ``A {{Variable Rate Coefficient Chlorine Decay Model}},''
  \emph{Environmental Science \& Technology}, vol.~43, no.~2, pp. 408--414,
  Jan. 2009.

\bibitem{elsherifControltheoreticModelingMultispecies2023}
S.~M. Elsherif, S.~Wang, A.~F. Taha, L.~Sela, M.~H. Giacomoni, and A.~A.
  Abokifa, ``Control-theoretic modeling of multi-species water quality dynamics
  in drinking water networks: {{Survey}}, methods, and test cases,''
  \emph{Annual Reviews in Control}, vol.~55, pp. 466--485, 2023.

\bibitem{moeiniBayesianOptimizationBooster2023}
M.~Moeini, L.~Sela, A.~F. Taha, and A.~A. Abokifa, ``Bayesian {{Optimization}}
  of {{Booster Disinfection Scheduling}} in {{Water Distribution Networks}},''
  \emph{Water Research}, vol. 242, p. 120117, Aug. 2023.

\bibitem{acrylamideNationalPrimaryDrinking2009}
{\relax OC}.~Acrylamide, ``National {{Primary Drinking Water Regulations}},''
  vol.~2, pp. 0--07, 2009.

\bibitem{engineeringconsultantImprovedWaterDistribution2012}
{Engineering Consultant}, W.~M. Grayman, S.~Kshirsagar, {Global Quality Corp.},
  M.~{Rivera-Sustache}, {USACE}, M.~Ginsberg, and {USACE}, ``An {{Improved
  Water Distribution System Chlorine Decay Model Using EPANET MSX}},''
  \emph{Journal of Water Management Modeling}, vol.~20, pp. R245--21, 2012.

\bibitem{tzatchkovAdvectionDispersionReactionModelingWater2002a}
V.~G. Tzatchkov, A.~A. Aldama, and F.~I. Arreguin,
  ``Advection-{{Dispersion-Reaction Modeling}} in {{Water Distribution
  Networks}},'' \emph{Journal of Water Resources Planning and Management}, vol.
  128, no.~5, pp. 334--342, Sep. 2002.

\bibitem{liImportanceDispersionNetwork2005}
Z.~Li, S.~G. Buchberger, and V.~Tzatchkov, ``Importance of {{Dispersion}} in
  {{Network Water Quality Modeling}},'' in \emph{Impacts of {{Global Climate
  Change}}}.\hskip 1em plus 0.5em minus 0.4em\relax Anchorage, Alaska, United
  States: American Society of Civil Engineers, Jul. 2005, pp. 1--12.

\bibitem{helblingModelingResidualChlorine2009a}
D.~E. Helbling and J.~M. VanBriesen, ``Modeling {{Residual Chlorine Response}}
  to a {{Microbial Contamination Event}} in {{Drinking Water Distribution
  Systems}},'' \emph{Journal of Environmental Engineering}, vol. 135, no.~10,
  pp. 918--927, Oct. 2009.

\bibitem{fisherComprehensiveBulkChlorine2017a}
I.~Fisher, G.~Kastl, and A.~Sathasivan, ``A comprehensive bulk chlorine decay
  model for simulating residuals in water distribution systems,'' \emph{Urban
  Water Journal}, vol.~14, no.~4, pp. 361--368, Apr. 2017.

\bibitem{boccelliOptimalSchedulingBooster1998}
D.~L. Boccelli, M.~E. Tryby, J.~G. Uber, L.~A. Rossman, M.~L. Zierolf, and
  M.~M. Polycarpou, ``Optimal scheduling of booster disinfection in water
  distribution systems,'' \emph{Journal of water resources planning and
  management}, vol. 124, no.~2, pp. 99--111, 1998.

\bibitem{liDisinfectantResidualStability2019a}
R.~A. Li, J.~A. McDonald, A.~Sathasivan, and S.~J. Khan, ``Disinfectant
  residual stability leading to disinfectant decay and by-product formation in
  drinking water distribution systems: {{A}} systematic review,'' \emph{Water
  Research}, vol. 153, pp. 335--348, Apr. 2019.

\bibitem{rossmanNumericalMethodsModeling1996}
L.~A. Rossman and P.~F. Boulos, ``Numerical {{Methods}} for {{Modeling Water
  Quality}} in {{Distribution Systems}}: {{A Comparison}},'' \emph{Journal of
  Water Resources Planning and Management}, vol. 122, no.~2, pp. 137--146, Mar.
  1996.

\bibitem{bashaEulerianLagrangianMethod2007a}
H.~A. Basha and L.~N. Malaeb, ``Eulerian--{{Lagrangian Method}} for
  {{Constituent Transport}} in {{Water Distribution Networks}},'' \emph{Journal
  of Hydraulic Engineering}, vol. 133, no.~10, pp. 1155--1166, Oct. 2007.

\bibitem{munavalliMULTISTEPEULERIANMETHOD2005}
G.~R. Munavalli and M.~S. Mohan~Kumar, ``{{MULTI-STEP EULERIAN METHOD FOR
  MULTICOMPONENT TRANSPORT IN WATER NETWORKS}},'' \emph{ISH Journal of
  Hydraulic Engineering}, vol.~11, no.~3, pp. 103--115, Jan. 2005.

\bibitem{blokkerImportanceDemandModelling2007}
M.~Blokker, J.~H.~G. Vreeburg, S.~G. Buchnberger, and J.~C. {van Dijk},
  ``Importance of demand modelling in network water quality models: A review,''
  in \emph{Conference {{High Quality Drinking Water}} 2007, {{Delft}}}.\hskip
  1em plus 0.5em minus 0.4em\relax Delft University of Technology, 2007, pp.
  1--10.

\bibitem{rossmanEPANETUserManual2020}
L.~A. Rossman, H.~Woo, M.~Tryby, F.~Shang, R.~Janke, and T.~Haxton,
  ``{{EPANET}} 2.2 {{User Manual Water Infrastructure Division}}, {{Center}}
  for {{Environmental Solutions}} and {{Emergency Response}},'' 2020.

\bibitem{abokifaInvestigatingImpactsWater2020a}
A.~A. Abokifa, L.~Xing, and L.~Sela, ``Investigating the {{Impacts}} of {{Water
  Conservation}} on {{Water Quality}} in {{Distribution Networks Using}} an
  {{Advection-Dispersion Transport Model}},'' \emph{Water}, vol.~12, no.~4, p.
  1033, Apr. 2020.

\bibitem{ozdemirDiscussionLagrangianMethod2023}
O.~N. Ozdemir, ``Discussion of ``{{Lagrangian Method}} to {{Model
  Advection-Dispersion-Reaction Transport}} in {{Drinking Water Pipe
  Networks}}'' by {{Feng Shang}}, {{Hyoungmin Woo}}, {{Jonathan B}}.
  {{Burkhardt}}, and {{Regan Murray}},'' \emph{Journal of Water Resources
  Planning and Management}, vol. 149, no.~1, p. 07022004, Jan. 2023.

\bibitem{fisherFrameworkOptimizingChlorine2018}
I.~Fisher, G.~Kastl, F.~Shang, and A.~Sathasivan, ``Framework for {{Optimizing
  Chlorine}} and {{Byproduct Concentrations}} in {{Drinking Water Distribution
  Systems}},'' \emph{Journal - American Water Works Association}, vol. 110,
  no.~11, pp. 38--49, Nov. 2018.

\bibitem{ding2024application}
Y.~Ding, Q.~Sun, Y.~Lin, Q.~Ping, N.~Peng, L.~Wang, and Y.~Li, ``Application of
  artificial intelligence in (waste) water disinfection: Emphasizing the
  regulation of disinfection by-products formation and residues prediction,''
  \emph{Water Research}, p. 121267, 2024.

\bibitem{constansUsingLinearPrograms2012}
S.~Constans, B.~Br{\'e}mond, and P.~Morel, ``Using {{Linear Programs}} to
  {{Optimize}} the {{Chlorine Concentrations}} in {{Water Distribution
  Networks}},'' pp. 1--12, Apr. 2012.

\bibitem{mala-jetmarovaLostOptimisationWater2017}
H.~{Mala-Jetmarova}, N.~Sultanova, and D.~Savic, ``Lost in optimisation of
  water distribution systems? {{A}} literature review of system operation,''
  \emph{Environmental Modelling \& Software}, vol.~93, pp. 209--254, Jul. 2017.

\bibitem{munavalliOptimalSchedulingMultiple2003}
G.~R. Munavalli and M.~S.~M. Kumar, ``Optimal {{Scheduling}} of {{Multiple
  Chlorine Sources}} in {{Water Distribution Systems}},'' \emph{Journal of
  Water Resources Planning and Management}, vol. 129, no.~6, pp. 493--504, Nov.
  2003.

\bibitem{trybyFacilityLocationModel2002}
M.~E. Tryby, D.~L. Boccelli, J.~G. Uber, and L.~A. Rossman, ``Facility location
  model for booster disinfection of water supply networks,'' \emph{Journal of
  Water Resources Planning and Management}, vol. 128, no.~5, pp. 322--333,
  2002.

\bibitem{pinedasandovalOptimalPlacementOperation2021a}
J.~D. Pineda~Sandoval, B.~M. Brentan, G.~M. Lima, D.~H. Cervantes, D.~A.
  Garc{\'i}a~Cervantes, H.~M. Ramos, X.~Delgado~Galv{\'a}n, and J.~D.~J.
  Mora~Rodr{\'i}guez, ``Optimal {{Placement}} and {{Operation}} of {{Chlorine
  Booster Stations}}: {{A Multi-Level Optimization Approach}},''
  \emph{Energies}, vol.~14, no.~18, p. 5806, Sep. 2021.

\bibitem{ayvazIdentificationBestBooster2015}
M.~T. Ayvaz and E.~Kentel, ``Identification of the {{Best Booster Station
  Network}} for a {{Water Distribution System}},'' \emph{Journal of Water
  Resources Planning and Management}, vol. 141, no.~5, p. 04014076, May 2015.

\bibitem{zhangOptimizingDisinfectantResidual2021}
C.~Zhang and J.~Lu, ``Optimizing disinfectant residual dosage in engineered
  water systems to minimize the overall health risks of opportunistic pathogens
  and disinfection by-products,'' \emph{Science of The Total Environment}, vol.
  770, p. 145356, May 2021.

\bibitem{prasadBoosterDisinfectionWater2004}
T.~D. Prasad, G.~A. Walters, and D.~A. Savic, ``Booster {{Disinfection}} of
  {{Water Supply Networks}}: {{Multiobjective Approach}},'' \emph{Journal of
  Water Resources Planning and Management}, vol. 130, no.~5, pp. 367--376, Sep.
  2004.

\bibitem{ardeshirControlTHMFormation2011}
A.~Ardeshir, M.~Alimohammadnezhad, K.~Behzadian, and F.~Jalilsani, ``Control of
  {{THM}} formation in multi-objective booster chlorination for water
  distribution systems,'' in \emph{Computing and {{Control}} for the {{Water
  Industry}}, {{CCWI Conference}}}, 2011.

\bibitem{behzadianNovelApproachWater2012}
K.~Behzadian, M.~Alimohammadnejad, A.~Ardeshir, F.~Jalilsani, and
  H.~Vasheghani, ``A novel approach for water quality management in water
  distribution systems by multi-objective booster chlorination,'' pp. 51--60,
  2012.

\bibitem{cozzolinoControlDBPsWater2005}
L.~Cozzolino, D.~Pianese, and F.~Pirozzi, ``Control of {{DBPs}} in water
  distribution systems through optimal chlorine dosage and disinfection station
  allocation,'' \emph{Desalination}, vol. 176, no. 1-3, pp. 113--125, Jun.
  2005.

\bibitem{maheshwariOptimizationDisinfectantDosage2020}
A.~Maheshwari, A.~Abokifa, R.~D. Gudi, and P.~Biswas, ``Optimization of
  disinfectant dosage for simultaneous control of lead and
  disinfection-byproducts in water distribution networks,'' \emph{Journal of
  Environmental Management}, vol. 276, p. 111186, Dec. 2020.

\bibitem{shangParticleBacktrackingAlgorithm2002}
F.~Shang, J.~G. Uber, and M.~M. Polycarpou, ``Particle {{Backtracking
  Algorithm}} for {{Water Distribution System Analysis}},'' \emph{Journal of
  Environmental Engineering}, vol. 128, no.~5, pp. 441--450, May 2002.

\bibitem{wangAdaptiveControlWater2005}
Z.~Wang, M.~M. Polycarpou, J.~G. Uber, and F.~Shang, ``Adaptive control of
  water quality in water distribution networks,'' \emph{IEEE transactions on
  control systems technology}, vol.~14, no.~1, pp. 149--156, 2005.

\bibitem{duzinkiewiczHierarchicalModelPredictive2005a}
K.~Duzinkiewicz, M.~Brdys, and T.~Chang, ``Hierarchical model predictive
  control of integrated quality and quantity in drinking water distribution
  systems,'' \emph{Urban Water Journal}, vol.~2, no.~2, pp. 125--137, Jun.
  2005.

\bibitem{wangHowEffectiveModel2021b}
S.~Wang, A.~F. Taha, and A.~A. Abokifa, ``How {{Effective}} is {{Model
  Predictive Control}} in {{Real}}-{{Time Water Quality Regulation}}?
  {{State}}-{{Space Modeling}} and {{Scalable Control}},'' \emph{Water
  Resources Research}, vol.~57, no.~5, p. e2020WR027771, 2021.

\bibitem{elsherifComprehensiveFrameworkControlling2024}
S.~M. Elsherif, A.~F. Taha, A.~A. Abokifa, and L.~Sela, ``Comprehensive
  {{Framework}} for {{Controlling Nonlinear Multispecies Water Quality
  Dynamics}},'' \emph{Journal of Water Resources Planning and Management}, vol.
  150, no.~2, p. 04023077, Feb. 2024.

\bibitem{leeMassDispersionIntermittent2004}
Y.~Lee, ``Mass dispersion in intermittent laminar flow,'' Ph.D. dissertation,
  University of Cincinnati, United States -- Ohio, 2004.

\bibitem{leaist1986absorption}
D.~G. Leaist, ``Absorption of chlorine into water,'' \emph{Journal of solution
  chemistry}, vol.~15, pp. 827--838, 1986.

\bibitem{hirschNumericalComputationInternal1990}
C.~Hirsch, ``Numerical computation of internal and external flows. {{Vol}}.
  2-{{Computational}} methods for inviscid and viscous flows,''
  \emph{Chichester, England and New York, John Wiley \& Sons, 1990, 708}, 1990.

\bibitem{durran2010numerical}
D.~R. Durran, \emph{Numerical methods for fluid dynamics: With applications to
  geophysics}.\hskip 1em plus 0.5em minus 0.4em\relax Springer Science \&
  Business Media, 2010, vol.~32.

\bibitem{kalmanMathematicalDescriptionLinear1963}
R.~E. Kalman, ``Mathematical description of linear dynamical systems,''
  \emph{Journal of the Society for Industrial and Applied Mathematics, Series
  A: Control}, vol.~1, no.~2, pp. 152--192, 1963.

\bibitem{gokhaleOptimizingControllabilityMetrics2021}
A.~Gokhale, S.~M. Valli, R.~Kalaimani, and R.~Pasumarthy, ``Optimizing
  controllability metrics for target controllability,'' in \emph{2021 {{Seventh
  Indian Control Conference}} ({{ICC}})}.\hskip 1em plus 0.5em minus
  0.4em\relax Mumbai, India: IEEE, Dec. 2021, pp. 141--146.

\bibitem{dattaNumericalMethodsLinear2004}
B.~Datta, \emph{Numerical {{Methods}} for {{Linear Control Systems}}}.\hskip
  1em plus 0.5em minus 0.4em\relax Academic Press, 2004.

\bibitem{wangTwoObjectiveDesignBenchmark2015}
Q.~Wang, M.~Guidolin, D.~Savic, and Z.~Kapelan, ``Two-{{Objective Design}} of
  {{Benchmark Problems}} of a {{Water Distribution System}} via {{MOEAs}}:
  {{Towards}} the {{Best-Known Approximation}} of the {{True Pareto Front}},''
  \emph{Journal of Water Resources Planning and Management}, vol. 141, no.~3,
  p. 04014060, Mar. 2015.

\bibitem{tsitsifliDisinfectionImpactsDrinking2018}
S.~Tsitsifli and V.~Kanakoudis, ``Disinfection {{Impacts}} to {{Drinking Water
  Safety}}---{{A Review}},'' \emph{Proceedings}, vol.~2, no.~11, p. 603, Jul.
  2018.

\end{thebibliography}

%
%

\end{document}